\documentclass[sigplan,10pt,nonacm]{acmart}
\renewcommand\footnotetextcopyrightpermission[1]{}
\usepackage[utf8]{inputenc}
\usepackage{enumitem}
\usepackage{tikz}
\usepackage{xcolor}
\usepackage{amsmath}
\usepackage{adjustbox}
\usepackage{pdfpages}
\usepackage{breakurl}
\usepackage{hyperref}
\usepackage{subfig}
\usepackage{filecontents}
\usepackage{graphicx}

\usepackage{tikz}
\usepackage{amsmath}

\newcommand{\note}[1]{{\bf{\textcolor{blue}{*** #1 ***}}}}
\newcommand{\mycomment}[1]{{}}
\newcommand{\name}[1]{{\em Parallax}}
\newcommand{\betree}[1]{{B$^{\epsilon}$--Tree}}
\newcommand{\betrees}[1]{{\betree{}s}}
\newcommand{\mmio}[1]{memory-mapped I/O}
\newcommand{\Mmio}[1]{Memory-mapped I/O}

\title{\Large \bf Balancing Garbage Collection vs I/O Amplification using hybrid Key-Value Placement in LSM-based Key-Value Stores}

\author{
{\rm Giorgos Xanthakis$^1$, Giorgos Saloustros, Nikos Batsaras$^1$,  Anastasios Papagiannis$^1$, Angelos Bilas$^1$}
Institute of Computer Science (ICS), Foundation for Research and Technology – Hellas (FORTH), Greece
100 N. Plastira Av., Vassilika Vouton, Heraklion, GR-70013, Greece\\
\{gxanth, gesalous,nikbats,apapag, bilas\}@ics.forth.gr} 

\begin{document}
\date{} 
\maketitle
\pagestyle{plain}

\footnotetext[1]{Also with the Department of Computer Science, University of Crete, Greece.}
\subsection*{Abstract}

Key-value (KV) separation is a technique that introduces randomness in
the I/O access patterns to reduce I/O amplification in LSM-based
key-value stores for fast storage devices (NVMe)~\cite{atlas, wisckey,
  tucana, kreon, hashkv, kvell, jungle, blobdb}. KV separation has a
significant drawback that makes it less attractive: Delete and
especially update operations that are important in modern
workloads~\cite{246158,258935} result in frequent and expensive
garbage collection (GC) in the value log.

In this paper, we design and implement \name{}, which proposes hybrid
KV placement that reduces GC overhead significantly and maximizes the
benefits of using a log. We first model the benefits of KV separation
for different KV pair sizes. We use this model to classify KV pairs in
three categories \emph{small, medium, and large}. Then, \name{} uses
different approaches for each KV category: It always places large
values in a log and small values in place.  For medium values it uses
a mixed strategy that combines the benefits of using a log and
eliminates GC overhead as follows: It places medium values in a log
for all but the last few (typically one or two) levels in the LSM
structure, where it performs a full compaction, merges values in
place, and reclaims log space without the need for GC.

We evaluate \name{} against RocksDB that places all values in place
and BlobDB~\cite{blobdb} that always performs KV
separation~\cite{246158}.
 We find that \name{} increases throughput by
up to 12.4x and 17.83x, decreases I/O amplification by up to 27.1x and 26x,
and increases CPU efficiency by up to 18.7x and 28x respectively, for
all but scan-based YCSB workloads.

\section{Introduction}
\label{sec:introduction}

Key-value stores typically use at their core the write-optimized
LSM-Tree~\cite{lsm} to handle bursty inserts and amortize write I/O
costs. LSM-Tree organizes data in multiple levels of increasing size.
Each data item travels through levels until it reaches the last level.
LSM-based designs have two important characteristics: 1) They always
produce large I/Os to the device and 2) They incur high I/O
amplification, up to several multiples of 10x compared to the dataset
size~\cite{spacerocksdb}.
 This is still the right tradeoff for hard disk drives (HDDs): Under
small, random I/O requests, HDD performance degrades by more than two
orders of magnitude, from 100s of MB/s to 100s of KB/s.  With the
emergence of fast block-based storage devices, such as NAND-Flash
solid state drives (SSDs) and block-based non-volatile memory devices
(NVMe), behavior is radically different under small, random I/Os: At
relatively high concurrency, these devices achieve a significant
percentage of their maximum throughput even with random I/Os.



Previous work has used a new technique, key-value (KV)
separation~\cite{atlas, wisckey, tucana, kreon, jungle, hashkv,blobdb}
to introduce some degree of randomness in I/Os generated by KV stores
and reduce I/O amplification. KV separation appends key-value pairs in
a value log as they are inserted (in unsorted order) and essentially
converts the KV store to a multistage index over the log. Therefore,
compaction operations across LSM levels involve only keys and
metadata, without moving values. This reduces I/O amplification
dramatically for large KV pairs.

\begin{figure}
\centering
\includegraphics[width=0.65\columnwidth]{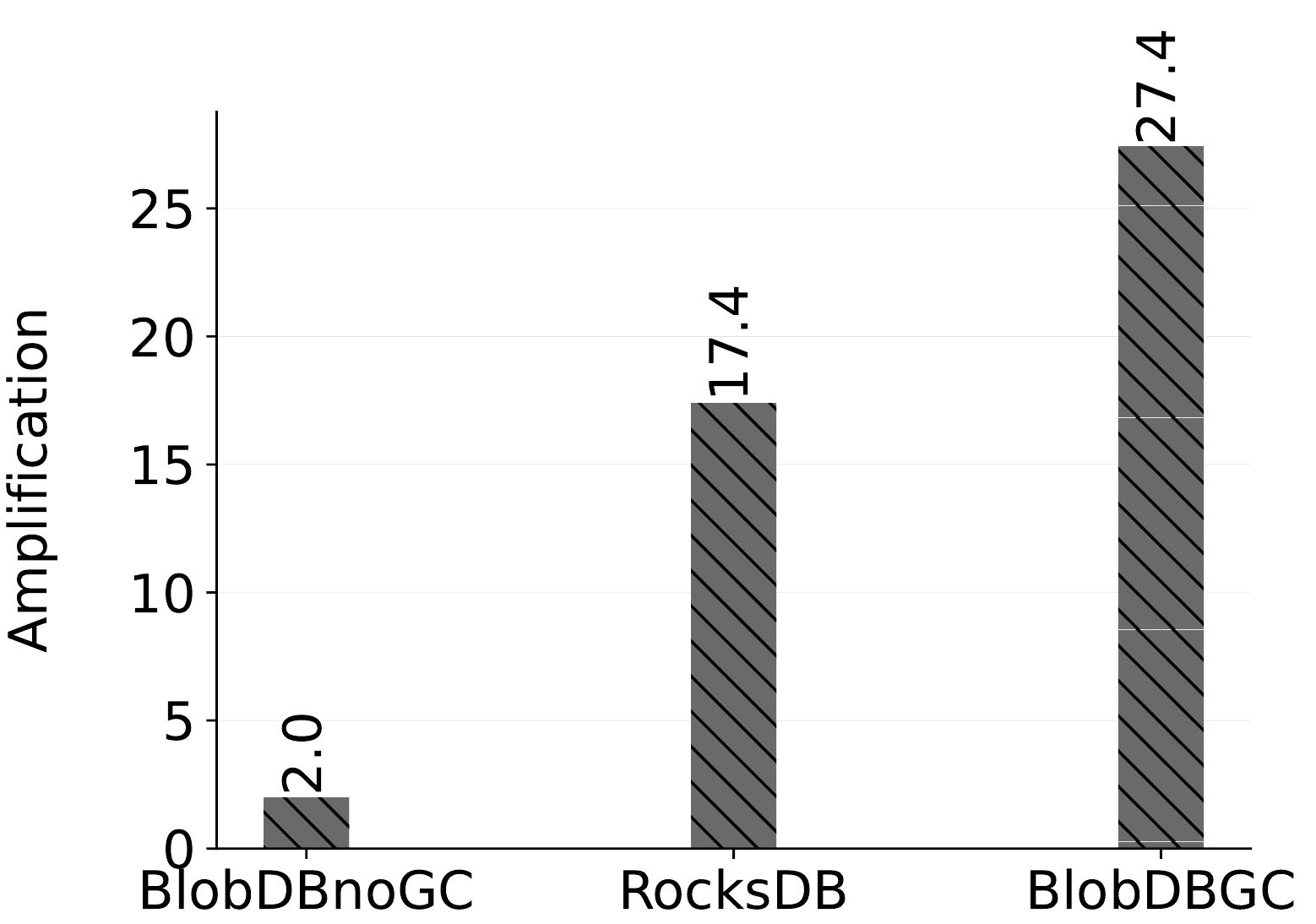}~
\caption{I/O amplification for BlobDB (with and without GC) and
  RocksDB for inserts of small (33 bytes) KV pairs.}
\label{fig:blobvsrocks}
\end{figure}

However, KV separation results in frequent garbage collection (GC) for
the log~\cite{hashkv,novkv}: Delete and especially update operations
that are common~\cite{246158,258935} generate old (garbage) values in the
value log that need to be garbage collected frequently to avoid
excessive space amplification.  Similar to past experience~\cite{lfs,
  seltzer2}, garbage collection in the value log is an expensive
process.
  Typically, GC requires two main and expensive operations: (1) Identify valid
values: We need to scan each log segment to identify if a value is the latest
value for a key in the dataset and therefore used. This requires a lookup read
for each value. Reads in the multistage index are already expensive in LSM-based
KV stores. This becomes exceedingly expensive as the number of keys in each log
segment increases, e.g. when there is pressure to free space eagerly or when the
KV pair size is small as shown in Figure~\ref{fig:blobvsrocks}. (2) Relocate
valid values: For values that are valid, they need to be copied to a new segment
at the end of the log, and metadata pointers that point to them need to be
updated, generating additional I/Os and high amplification. Then, the old
segment can be reclaimed for later use. Both operations (identify, relocate)
incur high overhead. Figure~\ref{fig:blobvsrocks} shows the effects of GC in I/O
amplification using RocksDB (no KV separation) and BlobDB (with KV separation)
for small KV pairs that dominate in Facebook production workloads
~\cite{246158}. There is a huge difference in BlobDB I/O amplification with and
without GC overhead, by more than 13x. When using GC, BlobDB I/O amplification
is even higher than RocksDB (27.4 vs. 17.4). Figure~\ref{fig:blobvsrocks} shows
only the identification cost of valid values since there are no deletes/updates
and no relocation occurs. The identification consumes valuable
read throughput from clients get/scan operations. In the case of
deletes/updates the I/O amplification increases even more since
the GC mechanism relocates valid KV pairs by appending them at the tail of the
log. The relocation operation consumes valuable write throughput from
inserts/updates, the Write Ahead Log, and compaction operations.

\mycomment{
  due to the fact that each read needs to search multiple
  levels. Therefore, garbage collection consumes a scarce resource
  which also translates to CPU cost for traversing the multistage
  index.  In addition, it may generate additional I/Os in cases where
  the bloom filters are not effective especially for the lower LSM
  levels that are large in size.  Relocating values requires copying
  the KV pairs in a new log segment, which although simple in nature,
  it consumes additional device read and write throughput. In
  addition, after moving a KV pair in the log, the respective pointers
  in the LSM metadata need to be updated, generating additional I/Os
  and high amplification.  }

\begin{figure}[t]
\subfloat[KV separation benefit.]{\includegraphics[width=0.5\columnwidth]{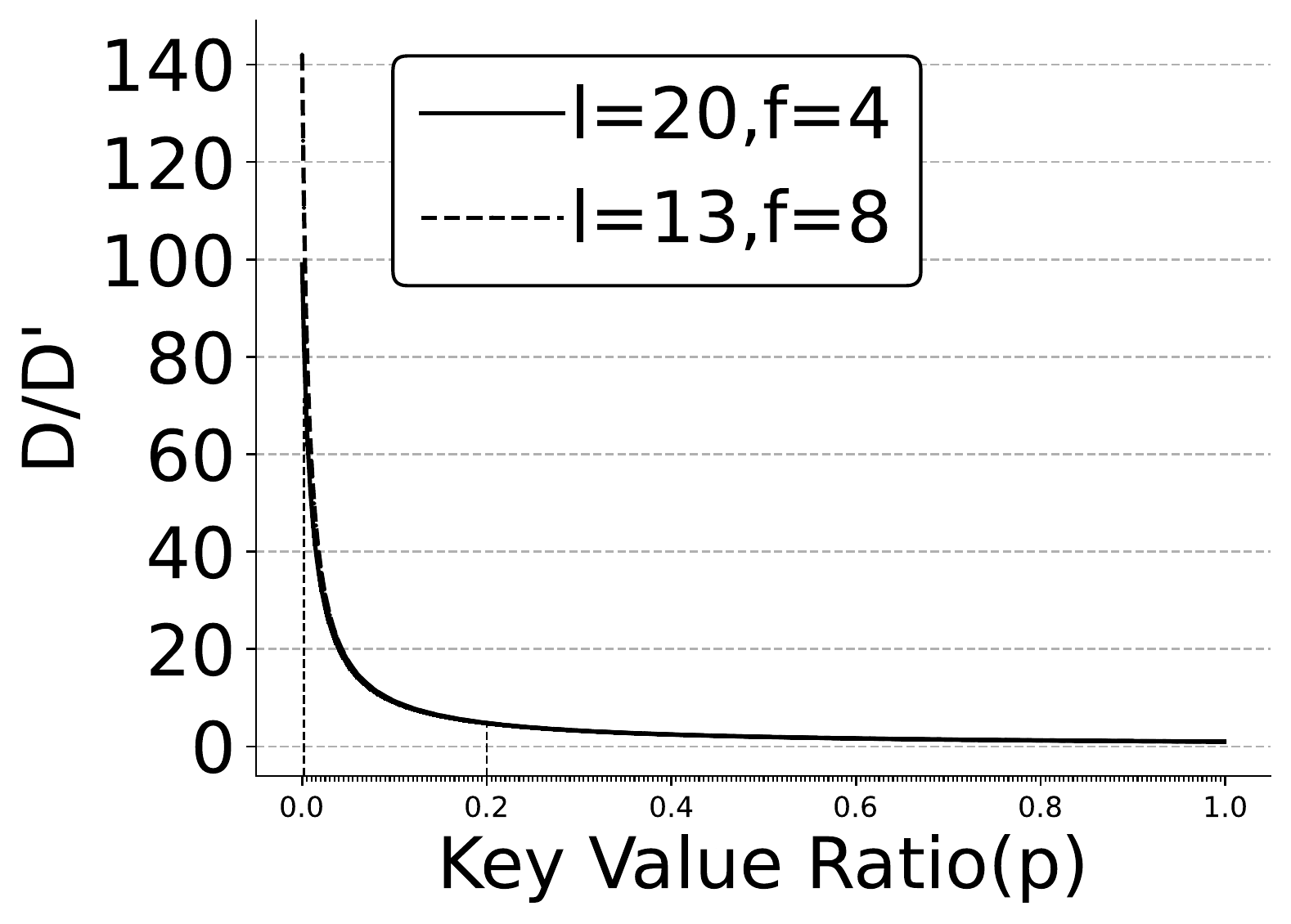}}
\subfloat[Capacity ratio of semi last levels over total.]{\includegraphics[width=0.5\columnwidth]{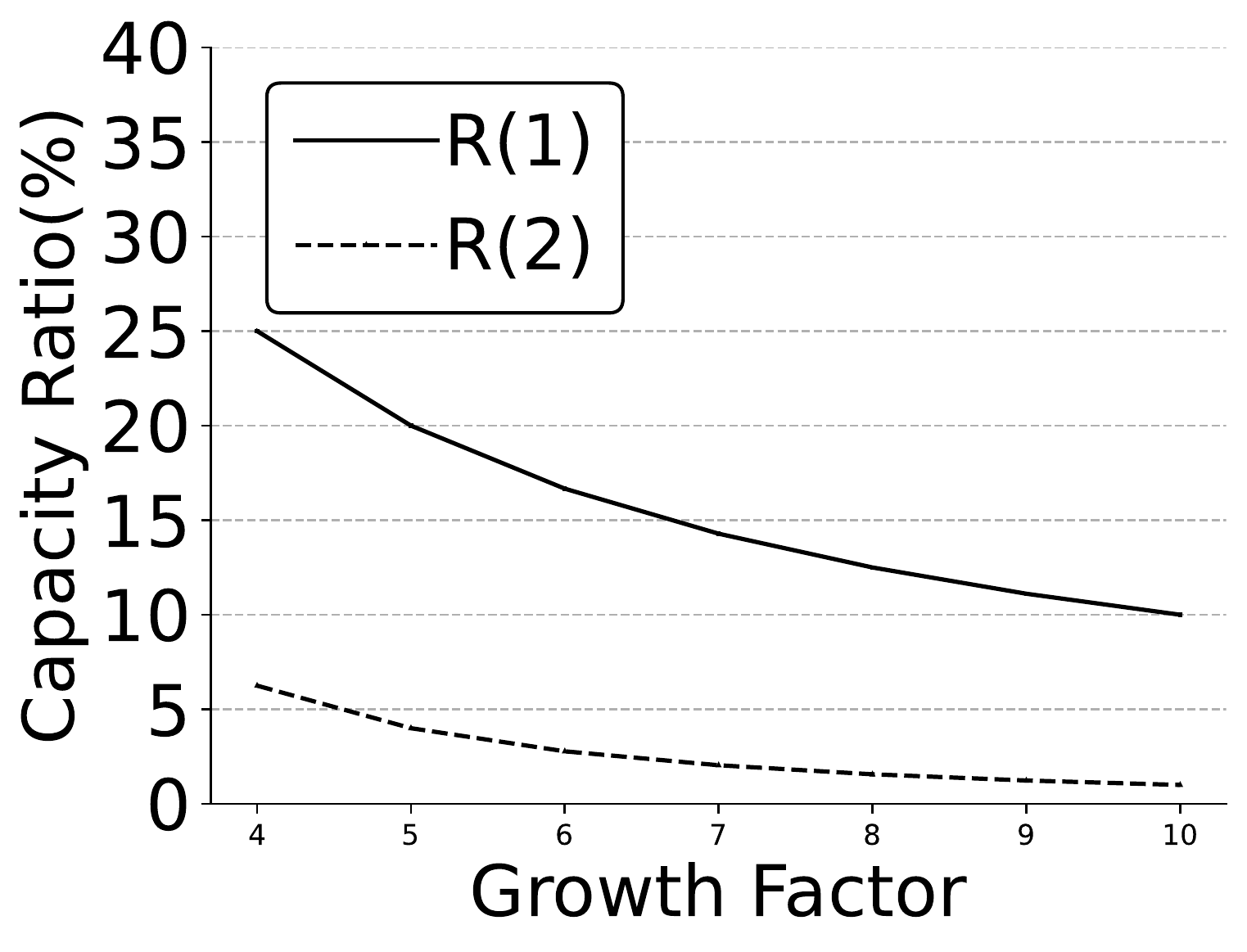}}
\caption{(a) Amplification ratio without and with KV separation
  ($\frac{D}{D'}$ in Equation~\ref{eq:logbenefit}) as a function of
  $p$ (x axis).  (b) Percentage of space occupied cumulatively by the
  first N-1,N-2, and N-3 levels compared to the total space occupied
  by the KV store for different growth factors.
}
\label{fig:valuelog}
\end{figure}

In this paper, we propose \name{}, an LSM-based KV store design for
fast storage devices that uses hybrid KV placement to address these
issues.  \name{} provides the benefits of using a log without the
excessive cost for GC overhead in the log, as follows. First, we use
the observation that workloads typical use KV pairs of different
sizes~\cite{246158}, including small, medium, and large KV pairs. In
particular, small KV pairs in many cases constitute a large percentage
(60\%) of the workload~\cite{246158}, although medium and large KV
pairs may dominate in terms of cumulative size. We model the benefits
of KV separation and we identify three size-based categories with
different behavior and benefits during KV separation
(Figure~\ref{fig:valuelog}): small KV pairs ($KV_{size} \leq 100$)
that do not benefit significantly in I/O amplification from using a log
($\leq 3x$), large KV pairs ($KV_{size} \geq 1024$) that
exhibit order-of-magnitude benefit (between $6x-12x$), and medium
 in between that have smaller but significant benefits (between
$3x-6x$).  In addition, we observe that large values incur low GC
overhead, small and medium values incur high GC overhead, with small
values introducing excessive GC costs.
\mycomment{
Our model shows that the benefits of KV separation are significant for
large values, and that it reduces I/O amplification by up to 12x
(Figure~\ref{fig:valuelog}).  On the other hand, for small KV pairs,
the benefit of KV separation is almost negligible. In addition, small
KV pairs introduce high GC overhead. For medium KV pairs, we show that
KV separation reduces amplification by up to 6x
(Figure~\ref{fig:valuelog}).  Although this benefit is smaller than
for large KV pairs, it is still significant.  However, medium size KV
pairs introduce high GC overhead as well, significantly deteriorating
the benefit of KV separation. 
}
This large variance in the benefits of KV separation
(Figure~\ref{fig:blobvsrocks}), combined with the significant overhead
of garbage collection in the value log makes KV separation less
attractive for workloads with mixed KV-pair sizes.

Based on our analysis and observations, we design \name{} that uses
different KV placement strategies for different KV pair sizes.
\name{} always places large KV pairs in a log with a clear benefit in
I/O amplification at low GC cost. \name{} places in the log large KV
pairs, even if they consist of small values and large keys.  \name{}
stores small KV pairs in place, within each LSM level. We use a
B+-tree index for each LSM level and store small KV pairs in its index
leaves, while it performs transfers from level to level as in LSM-type
approaches~\cite{lsm,blsm} (Figure~\ref{fig:blobvsrocks}).

For medium size KV pairs, \name{} uses a new technique: We place
medium KV pairs in a log up to the last level and then compact the
transient log in the last level, freeing the transient log. Given that the
transient log is freed when KV pairs are re-placed in the LSM structure,
there is no GC overhead associated for the transient log. Therefore,
medium KV pairs, combine most of the I/O amplification benefits with
almost no GC overhead.  To achieve this, \name{} essentially trades
space amplification for the transient log for a significant reduction in
I/O amplification. However, since all levels grow with a factor $f$,
typically 8 for space efficiency purposes~\cite{spacerocksdb}, the
space amplification in \name{} is limited.

Using hybrid KV placement in multiple logs and in-place introduces
challenges with ordering and recovery. \name{} uses log sequence
numbers to maintain ordering of keys within each region. In addition,
\name{} offers crash-consistency, and can recover to a previous (but
not necessarily the last) write, discarding all subsequent writes, as
is typical in modern KV stores~\cite{rocksdbwal, leveldbwal}.

We implement \name{} and evaluate a full-fledged \name{} prototype
with YCSB and different workloads. We compare \name{} to
RocksDB~\cite{rocksdb} that places all values in-place and with
BlobDB~\cite{blobdb} that uses KV separation. 
Our evaluation shows
that for YCSB workloads load A through run D with mixed size KV pairs,
\name{} compared to RocksDB and BlobDB 
increases throughput by
up to 12.4x and 17.83x, decreases I/O amplification by up to 27.1x and 26x,
and increases CPU efficiency by up to 18.7x and 28x
For range queries (run E-scans) with mixed size KV pair sizes \name{} 
has 7.95x more throughput than BlobDB and is 1.48x worse than RocksDB closing the gap 
compared to previous systems that perform KV separation~\cite{wisckey}.

Overall, the main contributions of our work are: 
\begin{enumerate}[noitemsep]
\item We propose hybrid KV placement that achieves most of the
  benefits of using a log for KV separation without excessive GC
  overhead.
\item We present an asymptotic analysis that describes I/O
  amplification in leveled LSM-based KV stores with and without KV
  separation and we use it to guide our design.
\item We design \name{} that provides hybrid KV placement, addressing
  issues of ordering and recovery, handling variable size keys and
  variable size updates for all logs and in-place values.
\end{enumerate}

\section{Modeling I/O Amplification}
\label{sec:vat}

In this section we start from an analytical model that calculates I/O
amplification in LSM key value stores which perform leveled
compaction~\cite{lsm}. Then, we calculate I/O amplification for KV
stores that use a value log and perform leveling
compaction~\cite{wisckey,kreon,hashkv}. Based on this analysis we
calculate the benefit for I/O amplification of placing KV pairs in a
log vs. in place, in the LSM levels themselves.

Amplification has two significant components: First, assuming level
size grows by $f$ times across consecutive levels, the system reads
and writes an excess of $f$ times more bytes, compared when merging
$L_i$ to $L_{i+1}$. Second, the cost of data reorganization across
multiple levels as data travel towards the lowest (largest) level: In
a system with $l$ levels, each data item moves through all levels
resulting in $l$ times excess traffic.  We refer to these quantities
of excess traffic as \textbf{\textit{merge amplification}} and
\textbf{\textit{level amplification}}, respectively.

Equation~\ref{eq:vatamp} captures I/O amplification in the insert path 
under the assumption that during a merge operation, the lower level is
fully read and written~\cite{lsm, blsm, wisckey, kreon, hashkv, rocksdb}. 

\begin{eqnarray}
D &= & \frac{S_l}{S_0}(S_0)  + 2\sum_{j=1}^{S_l/S_0}((j-1)\bmod{f})\cdot S_0\nonumber\\
& + & \frac{S_l}{S_1}(2S_1)  + 2\sum_{j=1}^{S_l/S_1}((j-1)\bmod{f})\cdot S_1\nonumber\\
& + & \ldots\nonumber\\
& + & \frac{S_l}{S_{l-1}}(2S_{l-1}) + 2\sum_{j=1}^{S_l/S_{l-1}}((j-1)\bmod{f})\cdot S_{l-1}
\label{eq:vatamp}
\end{eqnarray}

$D$ is the amount of I/O traffic produced until all $S_l$ data reach
$L_l.$ If $S_0$ is the size of the in-memory $L_0$ and $S_l$ is the
size of the last level, then we can assume that the entire dataset is
equal to $S_l$ and that all data will eventually move to the last
level $S_l.$ Then, $S_l/S_i$ is the total number of merge operations
from $L_i$ to $L_{i+1},$ until all data reach $L_l.$

Equation~(\ref{eq:vatamp}) consists of multiple sub-expressions
(rows), one per level, to capture level amplification. Each
subexpression captures merge amplification between two consecutive
levels, using two terms.

In each subexpression (row), the first term represents the data of the
upper (smaller) level that have to be read and written during the
merge operation. For each level $L_i$, $0 \leq i \leq l-1,$ each time
one of the $S_l / S_i$ merge operations occurs, all data stored in
$L_i$ are read and written, thus causing I/O traffic of size $2S_i$
(first term). Note that, in the first sub-expression for $L_0$ that
resides in memory, the factor of $2$ is missing in the first term,
indicating that we do not perform I/O to read data that are already in
memory.

The second term captures the total amount of data that are read and
written from $L_{i+1}$ in order to merge the overlapping ranges of
$L_i$ and $L_{i+1}$. The $\sum$ operator expresses the fact that
batches of size $S_i$ will require $S_l / S_i$ merge operations at the
corresponding level.  The $\bmod$ operator captures the fact that the
size of the lower (larger) level grows \emph{incrementally} up to $f$:
in the first merge operation the lower level has no data (i.e.,
$j-1=0$); in the next merge, the lower level contains data equal to 1x
the upper level; in each subsequent merge operation it contains data
2x, 3x, etc. of the data in the upper level. These data need to be
read and written during merging, hence the factor of $2$ before the
sum. We can re-write Equation~\ref{eq:vatamp} as:

\begin{equation}
\begin{aligned}[c]
Eq.~\ref{eq:vatamp} \Rightarrow
	D &= (2l-1)S_l + 2S_{l-1}\sum_{j=1}^{f^1}(j-1)\bmod{f} \\
	  &+ \ldots + 2S_0\sum_{j=1}^{f^l}(j-1)\bmod{f}\Rightarrow\\
	D &= (2l-1)S_l + 2S_{l-1}\Big(\frac{f^1}{f}\cdot \frac{(f-1)(f-1+1)}{2}\Big)\\
	  &+ \ldots + 2S_0\Big(\frac{f^l}{f}\cdot \frac{(f-1)(f-1+1)}{2}\Big)\Rightarrow\\
	D &= S_l(l -1 + fl)\label{eq:vatdata}
\end{aligned}
\end{equation}

Equation~\ref{eq:vatdata} expresses the amount of data read and
written during KV operation, until all data reach the lowest level.

\subsection{KV separation benefits}

Similarly to Equations~\ref{eq:vatamp} and~\ref{eq:vatdata} we
calculate traffic for KV stores that use KV separation. Each SST now
stores only keys and thus its size is equal to $K_i,$ with $S_i = K_i
+ V_i.$ The value log contains all KV pairs stored in the system, so
its size is $S_l.$ Consequently, we can write:

\begin{eqnarray}
D' &= &\frac{K_l}{K_0}(K_0)  + 2\sum_{j=1}^{K_l/K_0}((j-1)\bmod{f})\cdot K_0\nonumber\\
   &+ &\ldots\nonumber\\
   &+ &\frac{K_l}{K_{l-1}}(2K_{l-1}) + 2\sum_{j=1}^{K_l/K_{l-1}}((j-1)\bmod{f})\cdot K_{l-1}\nonumber\\
   &+ &S_l \Rightarrow\nonumber\\
D' &= &K_l\Big(l-1 + fl\Big) + S_l\label{eq:vatlogdata}
\end{eqnarray}

Equation~\ref{eq:vatlogdata} expresses the amount of data read and
written when using a KV log, until all data reach the lowest
level. The last term $S_l$ in Equation~\ref{eq:vatlogdata} represents
the fact that all KV pairs are appended once to the KV log.

Finally, the ratio $\frac{D}{D'}$ (Equations~\ref{eq:vatdata}
and~\ref{eq:vatlogdata}) expresses the benefit of KV separation over
in-place values.  If we assume that $p$ is the key to value size ratio
$p=K_l/(K_l+V_l)=K_l/S_l,$ then we can introduce $p$ in this ratio by dividing both
numerator and denominator with $S_l,$ which is the total size of
values. Therefore, we get:

\begin{equation}
\frac{D}{D'} = \frac{\frac{D}{S_l}}{\frac{D'}{S_l}} = \frac{(l-1+fl)}{p*(l-1+fl)+1}
\label{eq:logbenefit}
\end{equation}

\subsection{Discussion}

Figure~\ref{fig:valuelog}(a) plots this ratio as a function of $p.$
Based on this figure we can use two thresholds for p, $T_{SM}$ and
$T_{ML}$, to divide KV pairs in three categories, based on their
benefits in I/O amplification from KV separation:

\begin{enumerate}[noitemsep]
\item Large KV pairs with $0 < p \leq T_{ML},$ where the benefit can
  be more than an order of magnitude and where GC does not introduce
  significant overhead.
\item Small KV pairs with $T_{SM} < p \leq 1,$ where the benefit is
  small and the GC overhead becomes excessive.
\item Medium KV pairs, with $T_{ML} < p \leq T_{SM},$ where the
  benefit is smaller, but still substantial and will manifest only if
  we reduce the corresponding GC overhead.
\end{enumerate}


First, we should note that there are no models available for the cost
of GC in systems that use KV separation. Experimental evidence from
our and related work is that these overheads are high.  Delete and
especially update operations, which are common in modern
workloads~\cite{246158}, cause fragmentation in the value log. To
avoid high space amplification~\cite{spacerocks}, there is a need for
frequent garbage collection (GC) in the value log, which incurs high
overhead~\cite{seltzer2, hashkv, blobdb,novkv}.  Typically, the system
initiates GC periodically and after a configurable amount of update
(delete) operations. The log is usually organized as a list of
contiguous chunks of space (\emph{log segments}). GC scans KV pairs in
a configurable number of log segments to identify and relocate valid
KV pairs, using and updating the multilevel KV index.  Identifying
valid KV pairs incurs \emph{lookup cost}. Lookup cost depends
significantly on the number of KV pairs in a segment. Especially, for
small KV pairs this is high. In addition, lookup cost is independent
of the workload in that even with a small percentage of delete and
update operations GC must perform a lookup for each KV pair in a log
segment.  Relocating valid KV pairs at the end of the log incurs
\emph{cleanup cost} for transferring KV pairs and updating index
pointers to the new KV locations. Cleanup cost depends mostly on the
percentage of update and delete operations.

\mycomment{

It is important to note that lookups are already expensive operations
in LSM KV stores because they need to check multiple levels to
determine if a key exists. With the use of bloom filters that are
typical in such systems, in the best case each lookup requires at
least one large I/O operation for a single SST or two small I/O
operations in the case SSTs contain an index. With false positives for
bloom filters or when the bloom filters are not fully populated, this
can increase up to $l$ (number of levels) I/Os for each lookup.

}

Therefore, in our work we use $T_{ML}=0.02$ where the benefits of
placement in the log are so high that will not be offset by the
mediocre GC overhead of large KV pairs. On the other extreme, we use
$T_{SM}=0.2,$ beyond which point I/O amplification benefits are small
and will most likely be offset and exceeded by GC overheads for small
KV pairs. This leaves a relatively large range for medium KV pairs.
For instance, if keys are roughly 20 bytes, then small KV pairs have
values roughly below 80 bytes, large KV pairs have values larger than
roughly 1000 bytes, and medium KV pairs are in-between. We believe
that there is merit in examining these thresholds in more detail in
future work, taking into account other parameters as well, e.g. the
mix and percentage of different operation types (reads, inserts,
updates, deletes).

Second, we note that this model can be simplified for systems, such as
\name{}, where the index stores only fixed-size prefixes instead of
variable-size keys. To keep the model more general and applicable to
other systems as well, we use the above formulation. In practice, keys
are typically smaller than values and similar in size to prefixes.

Next, we present our design for \name{}.

\section{\name{} Design}
\label{sec:design}

The main idea in \name{} is to reduce GC overhead by using hybrid KV
placement, based on KV pair sizes: \name{} stores small KV pairs
always in-place, in the B+-tree leaves, and performs full compactions
for small KV pairs. For large KV pairs, \name{} always places KV pairs
in a dedicated log for large KV pairs and uses a dedicated GC process,
similar to previous work~\cite{blobdb, wisckey, kreon,hashkv}.  

\begin{figure*}[t]
\centering
\subfloat[\name{} Overview]{\includegraphics[width=2.1\columnwidth]{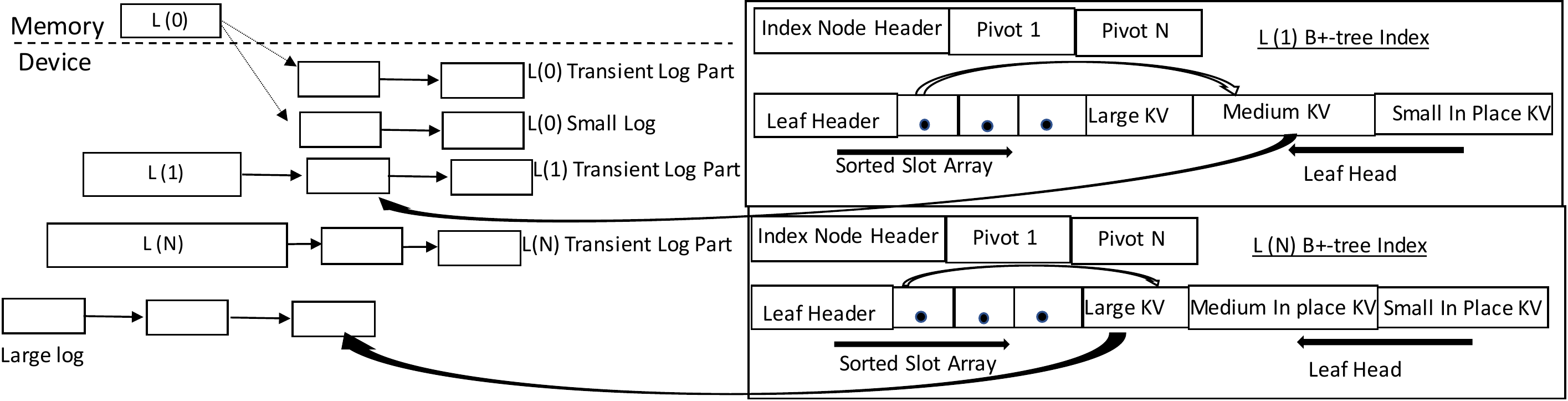}}
\\
\subfloat[Device Layout]{\includegraphics[width=1.5\columnwidth]{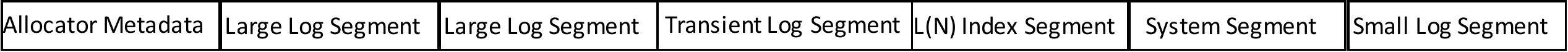}}
\caption{Overview of \name{} (a) index and log design and (b) device layout.}
\label{fig:parallax_arch2}
\end{figure*}

For medium KV pairs, \name{} uses a novel technique: It uses a
(transient) log to store KV pair during the first levels, e.g. up to
$L_i,$ and then merges the KV pairs in place for the remaining few
levels. When merging, \name{} stores medium KV pairs in the $L_i$
B+-tree index. Therefore, \name{} does not need to perform GC in the
transient log. Instead, it merely reclaims the log after compaction.
Although medium KV pairs are eventually merged in the LSM structure,
this technique results in significant benefits; in LSM-based KV stores
all levels, regardless of their size, contribute by the same
percentage to the overall I/O amplification since all KV pairs
traverse exactly the same path.

While placing KV pairs in the index and in multiple logs, \name{}
needs to deal with ordering and recovery of operations that modify
state (insert, update, delete).  Next, we first provide an overview of
\name{} and then we discuss how it handles KV pairs in each category.

\subsection{Overview}
\label{smlkvs}
Figure~\ref{fig:parallax_arch2}(a) shows an overview of \name{}.
\name{} is a leveled, LSM KV store that offers a dictionary API
(insert, delete, update, get, scan) of variable size KV pairs stored
in non-overlapping ranges, named \emph{\it{regions}}.  
KV regions share the same storage space through a common allocator~\cite{allocator},
as shown in Figure~\ref{fig:parallax_arch2}(b).
\name{} organizes each level as a full B+-tree~\cite{btree} index for
all KV pairs in the level~\cite{blsm, kreon}. For KV pairs that are
placed in logs it keeps a prefix of configurable size (12 bytes
currently) in the index.



\emph{Get} operations examine hierarchically all levels from $L_0$ to
$L_{N}$ and return the first occurrence.  \emph{Scan} operations
create one scanner per-level and use the index to fetch keys in sorted
order. They combine the results of each level to provide a global
sorted view of the keys.  To increase concurrency, each B+-tree index
implements Bayer's B+-tree concurrency
protocols~\cite{bayerconcurrency}. \emph{Delete} operations mark keys
with a tombstone and defer the delete operation similar to
RocksDB~\cite{rocksdb}, freeing up space at the next compaction.
Finally, \emph{update} operations are similar to a combined
\emph{insert} and \emph{delete}.

At each \emph{insert} operation, \name{} calculates the ratio $p$ of
prefix to KV pair size and uses $T_{SM}$ and $T_{ML}$ to categorize
each KV pair.
\name{} uses the prefix size as the nominator for $p$ and places KV
pairs in the log when the cumulative KV pair has a large size. In the
rest of this paper, and for simplicity we refer to as small, medium,
and large to describe the whole KV pair size. 
Based on the KV pair category (small, medium, large), \name{} uses the
respective mechanism for placement. It inserts in $L_0$ the
corresponding item or pointer to an item, in which case it also writes
the respective log for insert and update operations.

\subsection{Handling Small and Large KV pairs}

\name{} stores small KV pairs in the B+-tree index of each level as
follows. 
 Initially, \name{} inserts small KV pairs in the Small log for recovery purposes and then inserts them in its in memory $L_0$ B+-tree index.
The B+-tree index in each level consists of two types of
nodes: \emph{Index} and \emph{Leaf} nodes, as shown in
Figure~\ref{fig:parallax_arch2}(a). Index nodes store pivots, whereas
leaf nodes store either (a) a pointer to the KV location or (b) the
actual KV pair. For (a), \name{} uses also
prefixes~\cite{bohannon2001main} for the first \textit{M} bytes of the
key used for key comparisons inside a leaf. Prefixes reduce
significantly I/Os to the logs since leaves constitute the vast
majority of tree nodes~\cite{tucana,kreon,kvell}.
Index nodes and leaf nodes have a configurable size. In our case we
use 12~KB for index and 8~KB for leaf nodes respectively.

\name{} organizes its leaves dynamically
(Figure~\ref{fig:parallax_arch2}(a)), to store variable size KV pairs
or pointers to KV pairs: In each leaf, there are two dynamically
growing segments, the \emph{\it{slot array}} and the \emph{\it{data
  segment}}~\cite{modernbtreeteq,utree}. The slot array is a small
array where each cell is 4 bytes.  Each cell contains an offset inside
the leaf where the actual data are and grows from left to right. We
reserve the highest three bits of each cell in the array to store the
KV category. The data segment is an append only buffer that contains
pointers to the log or the in-place KV pairs and grows from right to
left. With this technique, \name{} is able store a dynamic number of
KV pairs per leaf because when the slot array and the data segment
borders interfere we know that the leaf is full. For update operations
we append the new value and update the slot array. When the data
segment runs out of space we decide either to compact the leaf if it
has fragmented space (due to updates) or to perform a typical split
leaf rebalance operation~\cite{btree}.

\name{} places large KV pairs in a log and uses a garbage collection(GC)
mechanism to reclaim free space. GC mechanism in \name{} works as follows. 
We use one dedicated GC thread for all regions, which we invoke
either synchronously when the system is under capacity pressure to reclaim space
or asynchronously, based on a condition.
\name{} keeps a private system region named \emph{\it{GC region}} where it keeps information about free space, similar to other systems~\cite{bdbjava}. 
This info includes the segments (set currently in 2~MB) of the large KV log
that have free space due to update/delete operations.

GC region consists only of small keys (16 byte KV size), so it only uses a Small log for recovery purposes and has a small memory/space footprint.
For a device of 2~TB  capacity, 2~MB segments, and only large KVs, the GC region size in the worst case is 16~MB.
When compaction threads discover a deleted/updated for a large KV, they 1) use the KV's device offset to locate the corresponding large log segment start offset. Since all space in \name{} is segment aligned, compaction threads calculate the segment start offset through a modulo operation 2) update segment free space counter in the GC region using segment's start offset as key.
The GC thread wakes up periodically to check the state of the free space. If a segment's free space exceeds a preconfigured threshold (10\%), it performs the following steps.
First, it iterates over all the segment KV pairs and issues look-up operations to the multilevel index to see which KVs are valid.
Finally, it transfers the valid ones via a put operation to the corresponding region and reclaims the segment.


\subsection{Handling Medium KV Pairs}
\label{subsec:medium_log}

For medium KV pairs, \name{} uses a transient log to reduce I/O amplification
and merges the log in place to reclaim the full log space without the
need for GC.
%
Using a transient log for medium KV pairs, raises two questions:
(a) What is the size of the transient log and the
associated space amplification?  (b) What is the cost of merging the
transient log back in the LSM structure?

\paragraph{Transient log size:}
We notice that a value log does not grow significantly in size
for the first levels in the LSM-tree. The cumulative capacity of the
first levels in an LSM-based KV store is a small percentage of the
last one or two levels. Given a growth factor $f$, we can calculate
the total capacity $S_N$ of $N$ levels as $S = S_0 *
\frac{(1-f^N)}{(1-f)}.$ Similarly, the aggregate capacity $S_{N-1}$ of
the N-i first levels is $S_{N-i} = S_0 * \frac{(1-f^{N-i})}{(1-f)}.$
Then, we can calculate the ratio $R(i)$ of capacity for the first N-i
levels compared to N levels as $R(i)=\frac{1-f^{N-i}}{1-f^N}.$

Figure~\ref{fig:valuelog}(b) shows R(1) and R(2) 
for growth factors between 4 and
10. We use growth factor from 4 to 10 because growth factor 4 is
optimal for the LSM-tree and results in minimum I/O amplification.
Larger growth factors in the range 8-10, increase I/O amplification
slightly in favor of lower space amplification for workloads with high
update ratios. If we assume that the full dataset is placed at the
last level and all intermediate levels are essentially updates, then a
smaller growth factor results in relatively more space for
intermediate LSM levels, increasing space amplification compared to
the dataset size (last level).  Therefore, production systems prefer
to use growth factors around 8-10.

We see that in the worst-case scenario, where all KV pairs belong to
the medium category, using a log up to, but excluding, the last level
$L_{N-1}$
(R(1)) will delay freeing between 10\% ($f = 8$) and 25\% ($f=4$) of the
device capacity until we merge values back to the LSM-tree. At the
same time, we get almost all benefit of using a log, by not
reorganizing medium KV pairs for all but the last level. If we merge
medium KV pairs at level $L_{N-2}$ (R(2)), then we will delay freeing
at most 6\% of the space. In both cases, all space will be freed as
medium values are merged to the last one or two levels.

\begin{figure}[t]
\centering
\includegraphics[width=\columnwidth]{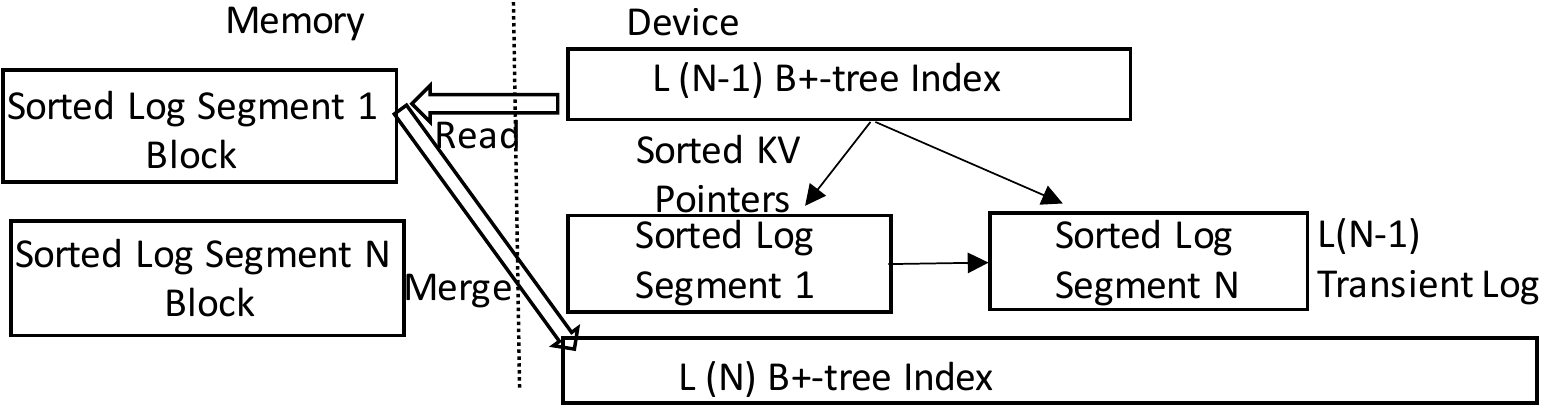}
\caption{Transient log compaction process at the $L_{N-1}$ level.}
\label{fig:medium_log_compaction}
\end{figure}

\paragraph{Merge cost for the transient log:}
A basic prerequisite of the compaction process in the
LSM-tree~\cite{lsm} is to insert keys from $L_{i-1}$ to $L_i$ in
sorted order to amortize I/O costs.  Otherwise, this process does not
amortize I/O costs and just performs redundant data transfers. As a
result, we must insert the KV pairs of the transient log in sorted order.

The index of $L_{N-1}$ already contains the pointers to the KV pairs
of the transient log sorted. However, a full scan of the transient log in
this case causes a significant penalty in traffic for the following
reason. Medium KV pairs are in the order of hundred of bytes compared
with the minimum block size (4 KB) of the device and are in random
order. As a result the system may end up performing one 4 KB I/O
operation for a few hundred bytes resulting in high I/O amplification,
e.g. up to 40x the size of the transient log for 100 byte KV
pairs.

To overcome this cost, \name{} first appends medium KV pairs in the Small log along with small KVs for recovery purposes.
It is important to notice that the Small log in \name{} has the equivalent role of a Write-Ahead-Log.
Then, it fully stores medium KV pairs in memory in $L_0$. 
During compaction from $L_{0}$ to
$L_1,$ \name{} uses its $L_0$ B+-tree
index to insert in $L_1$ KV pairs in a sorted manner. Specifically,
it appends the medium KV pairs in the transient log and inserts the pair <prefix,pointer> in the $L_1$ index.


Then, each segment of the transient log is attached to a particular LSM
level and travels down the LSM hierarchy with compaction operations
alongside the corresponding index
(Figure~\ref{fig:parallax_arch2}(a)). When a transient log segment
reaches the last level, it can be reclaimed as a whole after merging
its contents back to the LSM structure.

During merging of each transient log segment, \name{} needs to fetch
the segment once and incrementally, as shown in Figure~\ref{fig:medium_log_compaction}.
 For example, in the case of an $L_{N-1}$ size of 200 GB and
growth factor 8 (so $L_N$ = 1.6~TB) and transient log segments of
8~MB, \name{} needs about 200~MB of memory to perform the merge
operation, if it fetches 8KB at a time from each segment.  It is
important to note that this process works because fast storage
devices, such as NVMe, allow us to perform 8~KB random read I/Os at
high throughput (at approximately 80\% of the optimal device
throughput in our case).

Finally, to ensure that each compaction satisfies the growth factor we
keep two sizes for each level $L_i$ using as size for medium KV pairs:
1) The size of the prefix + log pointer and 2) their actual key +
value size. We use the former as the size for $L_i$ when merging to it
the previous level $L_{i-1}$ and the later as the size for $L_i$ when
merging it to the next level $L_{i+1}$.

\subsection{Space Management, I/O Paths, and Recovery}
In \name{}, each region consists of a per-level B+-tree index and the
three KV-pair logs, as shown in Figure~\ref{fig:parallax_arch2}(a).
Each of these entities allocate space at a large (segment) granularity
(currently 2~MB in \name{}), as shown in
Figure~\ref{fig:parallax_arch2}(b).
Writes in \name{}
occur either for writing a chunk (256~KB) of a log segment or during compaction to write the new merged level 
in segment (2~MB) granularity.
On the other hand, get and scan operations in \name{} generate by design (logs and B+-tree index) small (4~KB) and random I/Os to reduce amplification~\cite{tucana,kreon}.
Finally, \name{} uses direct I/O in segment granularity to read levels $L_i$ and $L_{i+1}$ during compaction.

During initialization, \name{} maps to its address
space the available storage, either from block devices or files.
\name{} uses two different I/O paths~\cite{lmdb}:
\begin{enumerate}
\item Memory-mapped I/O to read data from logs and traverse each level index during get and scan operations.  
\item Direct I/O (with write() and read() system calls)  for writing all logs and read/write for merging levels during compaction.
\end{enumerate}
Memory-mapped I/O is a good fit for \name{}'s read access pattern (small and random). The reasons for this are 1) It saves CPU cycles (up to 30\%)~\cite{kreon,harizopoulos,aquila} compared to a userspace cache, especially when the data are already in DRAM, and 2) It avoids copies from kernel to userspace. In particular, \name{} uses Fastmap~\cite{fastmap-github,fastmap} open-source project
which is a custom mmap I/O path optimized for storage. Fastmap also provides the ability to set the memory size; it statically allocates the configured memory on initialization.
On the contrary, \name{} avoids memory-mapped I/O for writes and uses direct I/O for the following reasons:
\begin{enumerate}
\item Write requests are always large (order of hundreds of~KB); thus, consecutive 4~KB write page faults (instead of a system call) to issue large write I/Os adds CPU overhead.
\item Lack of direct control to issue the write I/O requests - msync() blocks all page faults in the process due to the root (per process) page table lock~\cite{fastmap}.
\item Write operations from compaction and log through mmap pollute the cache. 
\end{enumerate}

Compactions read the levels via read() system calls in segment (2~MB) granularity and extract KVs from leaves that are already in sorted order per level. Then they merge-sort the levels into the new $L'_{i+1}$ and build its B+-tree index bottom-up since KVs arrive in sorted order. As a result, B+-tree leaves for $levels\geq 1$ are always full. Furthermore, \name{} saves CPU cycles since insert operations
do not traverse the tree levels; they append the next KV in the current $L_{i+1}'$ leaf.
For each region's log \name{} keeps a circular \emph{\it{tail buffer}} where it appends new entries. When a chunk of the log buffer is full (256~KB), it appends it to the device. Get and Scan operations from $L_0$ that dereference log pointers (Medium, Large) check if the pointer resides in memory or the device.

On compaction completion, \name{} frees the space of $L_i$ and $L_{i+1}$ and especially for $L_0$ to $L_1$ compactions reclaim part of the Small log which contains the compacted KV pairs. Then,
 \name{} logs in a global (for all regions) redo-log three vital info for recovery 1) The list of segments allocated for $L_{i+1}'$ tree,
2) The list of freed segments, and
3) The entry that describes the new level in the system catalog. Specifically, for $L_0$ to $L_1$ compactions it records also the offset of the logs up to which \name{} has added entries to $L_1$. Upon recovery, \name{}
replays this log applies the changes to its in-memory allocator metadata, updates its system catalog, and replays its logs (Small, Large) to recreate $L_0$. \name{} periodically persists its catalog and allocator metadata info to reduce recovery time. It is important to notice that except for the Small log, which contains small and medium KVs of $L_0$ Large log serves recovery purposes. 

\name{} needs to also deal with KV pairs that change category after an
update that may increase or reduce its size.  To solve this issue and
to maintain ordering of operations within each region, we use a Log
Sequence Number (LSN) per log entry. LSN is an eight-byte, per-region
counter which we atomically increment before appending to any of the
logs and we store it with each log entry.  During region recovery,
\name{} replays each log entry of the three logs with the correct
order, as indicated by their LSN number.

Finally, \name{}, as most other KV stores~\cite{rocksdbwal,
  leveldbwal}, acknowledges writes as soon as they are written in
memory after the group commit (flush). Therefore, \name{} can recover
to a previous consistent point, which may not include the
last (acknowledged) write. Most KV stores (and \name{}) can be
configured to acknowledge writes after they are written to the device
or to perform more frequent flush operations, but these are not
commonly used as they increase acknowledgment delay or I/O overhead.


\section{Methodology}
\label{sec:method}

Our testbed consists of a single server which runs the key-value store
and the YCSB~\cite{ycsb} client. The server is equipped with two
Intel(R) Xeon(R) CPU E5-2630 running at 3.2~GHz, with 12 physical
cores for a total of 24 hyper-threads and with 256~GB of DDR4 DRAM. It
runs CentOS 7.3 with Linux kernel 4.4.159. The server has one Intel
Optane P4800X device model with total capacity of
375~GB~\cite{intel4}.

\begin{table}[]
\centering
\resizebox{0.99\columnwidth}{!}{
\begin{tabular}{|l|c|c|c|c|}
\hline
Workload 
& \begin{tabular}[c]{@{}c@{}}KV Size Mix\\ S\%-M\%-L\%\end{tabular} 
& \begin{tabular}[c]{@{}c@{}}\# KV pairs \\ (Millions) \end{tabular} 
& \begin{tabular}[c]{@{}c@{}}Cache \\ Size (GB)\end{tabular} 
& \begin{tabular}[c]{@{}c@{}}Dataset\\ Size (GB)\end{tabular} \\ \hline
Small (S) & 100-0-0 & 500 & 2 & 10 \\ \hline
Medium (M) & 0-100-0 & 200 & 4 & 26 \\ \hline
Large (L) & 0-0-100 & 100 & 16 & 100 \\ \hline
\begin{tabular}[c]{@{}l@{}}Small\\ Dominated (SD)\end{tabular}  & 60-20-20 & 100 & 4 & 22.5 \\ \hline
\begin{tabular}[c]{@{}l@{}}Medium\\ Dominated (MD)\end{tabular}  & 20-60-20 & 100 & 4 & 25.5 \\ \hline
\begin{tabular}[c]{@{}l@{}}Large\\ Dominated (LD)\end{tabular}  & 20-20-60 & 100 & 4 & 62.5 \\ \hline
\end{tabular}
}
\caption{Workloads description in number of KV pairs, cache size, and
  dataset size. Small KV pairs are up to 119 Bytes, Medium KV
  pairs have sizes between 120-1023 Bytes, and Large KV pairs have
  size greater than 1024 Bytes. 
}

\label{tab:workloads}
\end{table}

We compare \name{} to RocksDB that places all KV pairs in-place and to
BlobDB that always performs KV separation.  We configure
BlobDB and RocksDB v6.11.4 with 128MB for the WAL and 4 threads for background I/O operations (log flushing and compactions),
on top of \textit{XFS} with disabled compression and
jemalloc~\cite{jemalloc}, as recommended.  We configure RocksDB to use
direct I/O because we evaluate experimentally that in our testbed
results in better performance. Furthermore, we use RocksDB's
user-space LRU cache, by varying the size of the cache based on the
workload which we similarly use for \name{} as shown in
Table~\ref{tab:workloads}. To have an equal comparison we disable
bloom filters for RocksDB and \name{} as BlobDB does not yet support
them.  For BlobDB we set GC to scan 30\% of the log when the GC threads wake up after a compaction.
For \name{} we set GC to reclaim a log segment when 10\% of the segment is invalid.
We choose these thresholds for the GC mechanism on each system because preventing space waste is a high priority in production~\cite{XengineAlibaba}.

We evaluate six workloads in terms of KV sizes, three that use KV pairs with a
single size (Small, Medium, Large) and three that use mixed KV pair sizes
(Small-Dominated, Medium-Dominated, and Large-Dominated), as suggested by
Facebook and Twitter workloads~\cite{246158,memcachedworkloads,258935}. Each
workload can be described as the percentages of the KV pair sizes it includes,
e.g. 100\% small KV pairs. KV pair sizes are categorized as small, medium, or
large, based on the analysis of Section~\ref{sec:vat}. Table~\ref{tab:workloads}
summarizes these workloads. 
Additionally, the last two columns in
Table~\ref{tab:workloads} describe the dataset size for each workload and the cache size used for all systems for each workload.

We use the following key and value sizes for each category. All
categories have a key size of 24 bytes on average. The value size is 9
bytes for small KV pairs, 104 bytes for medium KV pairs, and 1004
(YCSB) for large KV pairs.  These result in $p=0.02$ for large KV
pairs, $p=0.72 > 0.2$ for small KV pairs, and $p=0.19$ (between 0.02
and 0.2) for medium KV pairs, closer to the small rather than the
large category as this is more representative of actual workloads.


We also vary the type of operations, based on YCSB. We use Load A and Run A for
large parts of our analysis as the mixes of these two operations exhibit the lookup and
cleanup costs of our analysis of garbage collection in the log: Load A includes
100\% insert operations and exhibits only the lookup cost of GC, whereas Run A
includes update (50\%) and read (50\%) operations and exhibits both lookup and
cleanup costs. 
Furthermore, in the case of mixed workloads, update operations of Run A change the sizes of KV pairs and thus their category.
We also pay attention to Run E
that includes scan operations which are important for systems that use KV
separation. Finally, we also show results for the full YCSB workloads.

In all cases, we examine throughput in Kops/sec, I/O amplification as
ratio of device traffic (reads and writes) over application traffic,
and efficiency in Kcycles/op. We calculate cycles/op as:

$$cycles/op = \frac{\frac{CPU\_utilization}{100} \times \frac{cycles}{s} \times cores}{\frac{average\_ops}{s}},$$

where $CPU\_utilization$ is the average CPU utilization for all CPUs,
excluding idle and I/O wait time, as given by~\textit{mpstat}. As
$cycles/s$ we use the per-core clock frequency, $average\_ops/s$ is
the throughput reported by YCSB, and $cores$ is the number of system
cores including hyperthreads.

We use a C++ version of YCSB~\cite{Jinglei2016} and we
modify it to produce different values according to the KV pair size
distribution we study.  In all systems we use a total of 8 databases
and 16 threads respectively.

Finally, we use $T_{SM}=0.2$ and $T_{ML}=0.02$ and
we explore two variants.  We leave a more detailed exploration of
$T_{SM}$ and $T_{ML}$ based on the workload in terms of KV size and
operation distributions, for future work.

\begin{figure*}[t]
\centering
\subfloat[Throughput]{\includegraphics[width=.34\textwidth]{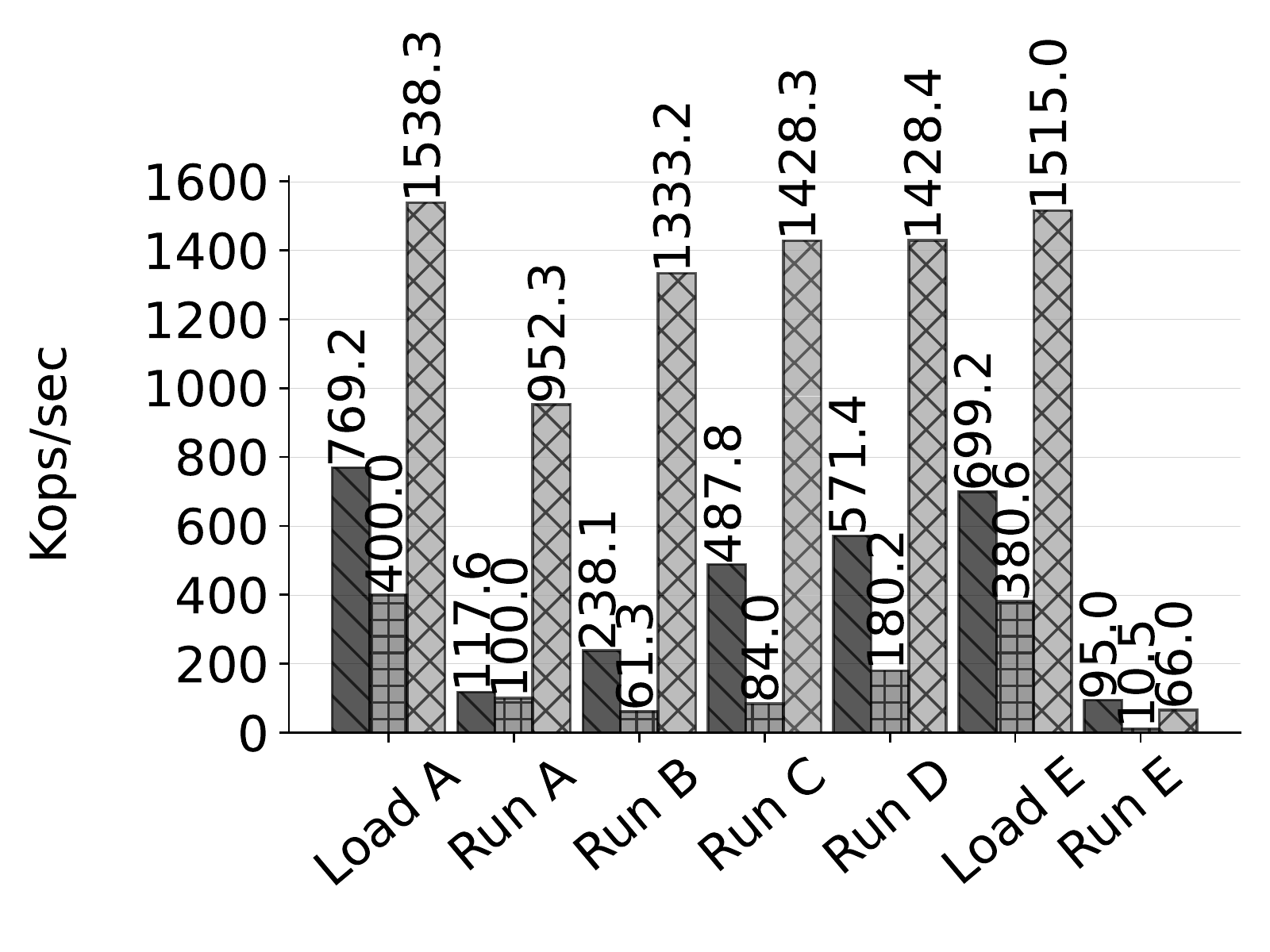}}
\subfloat[I/O Amplification]{\includegraphics[width=.34\textwidth]{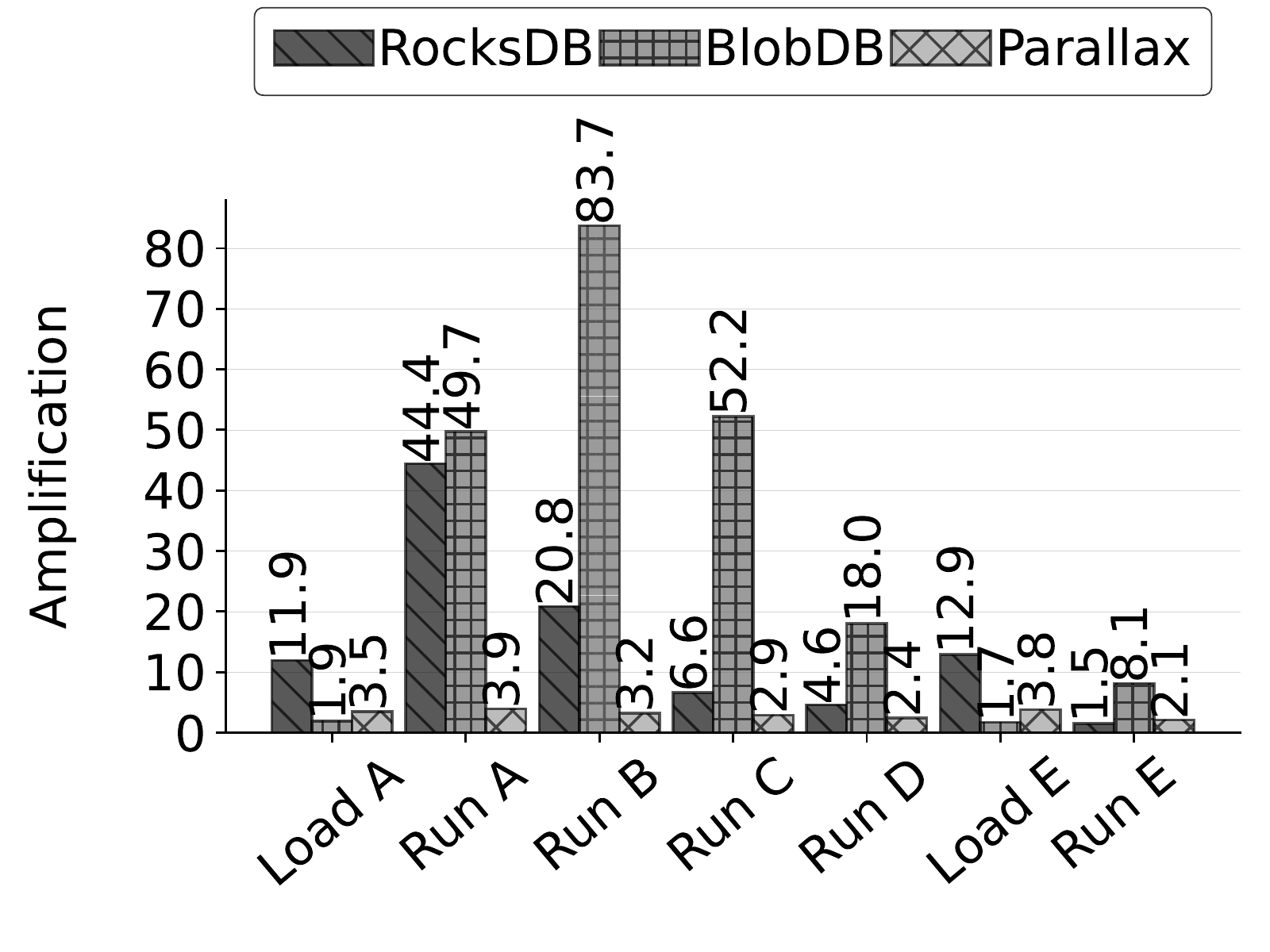}}
\subfloat[Efficiency]{\includegraphics[width=.34\textwidth]{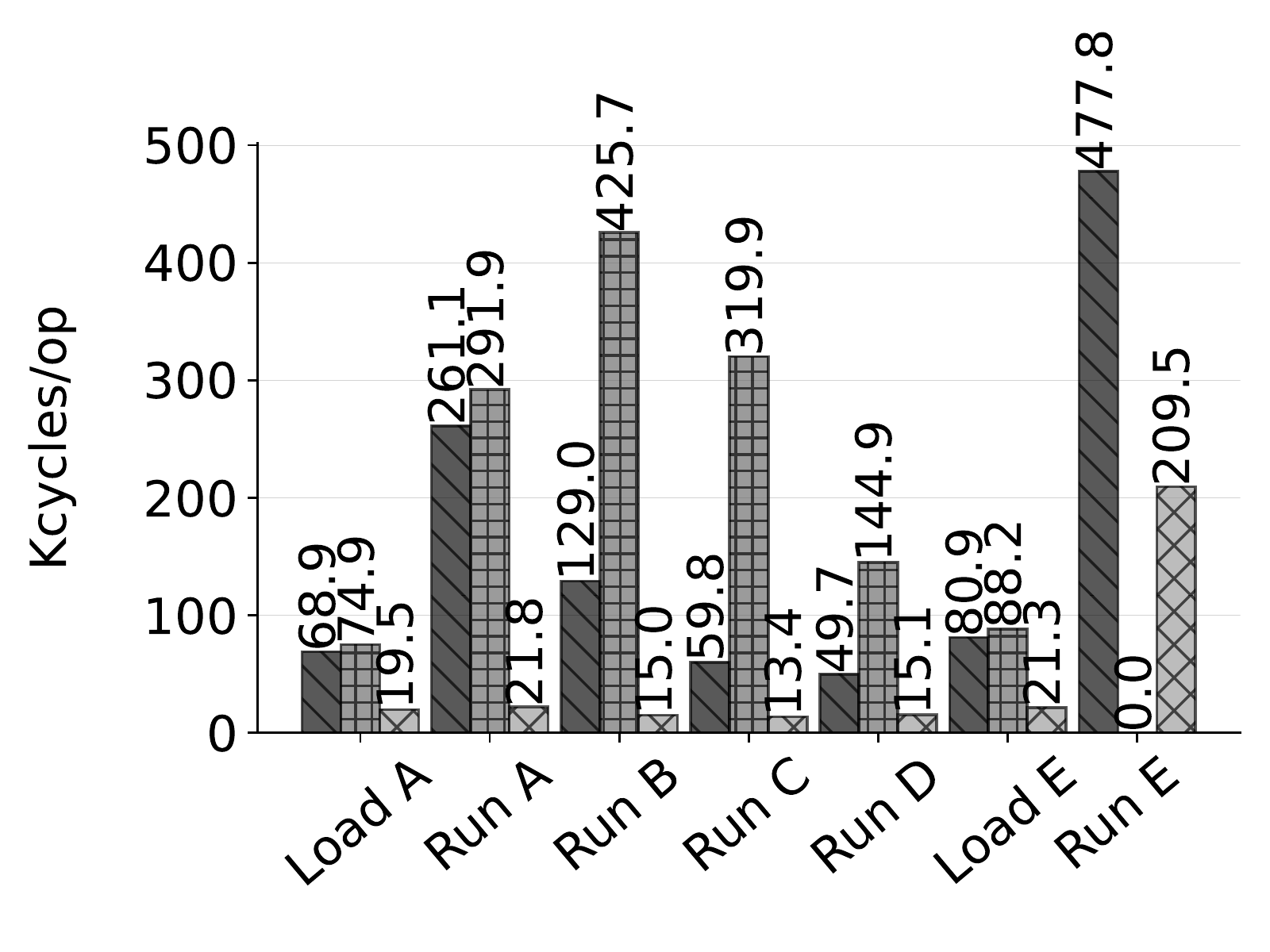}}
\\
\subfloat[Throughput]{\includegraphics[width=.34\textwidth]{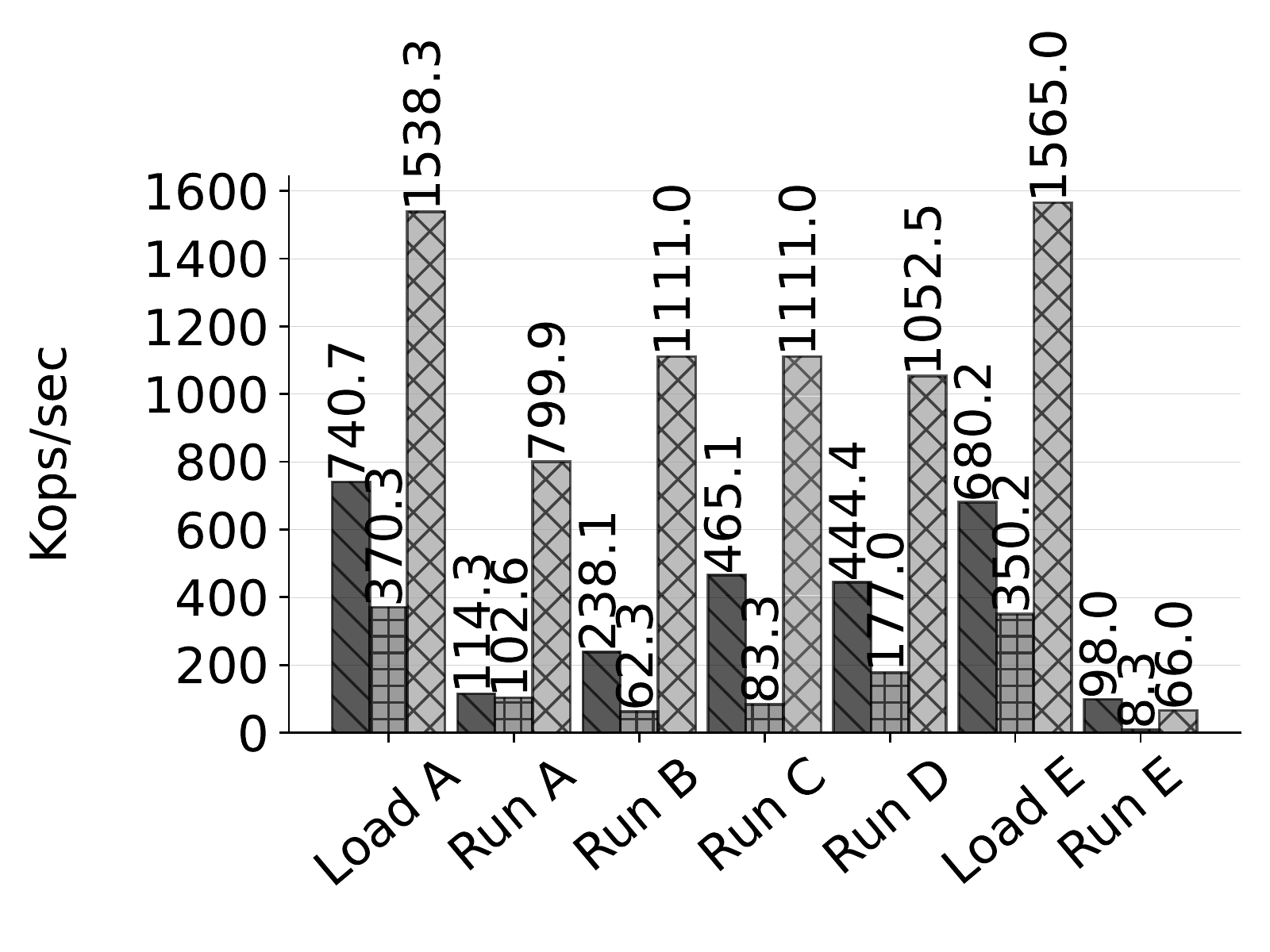}}
\subfloat[I/O Amplification]{\includegraphics[width=.34\textwidth]{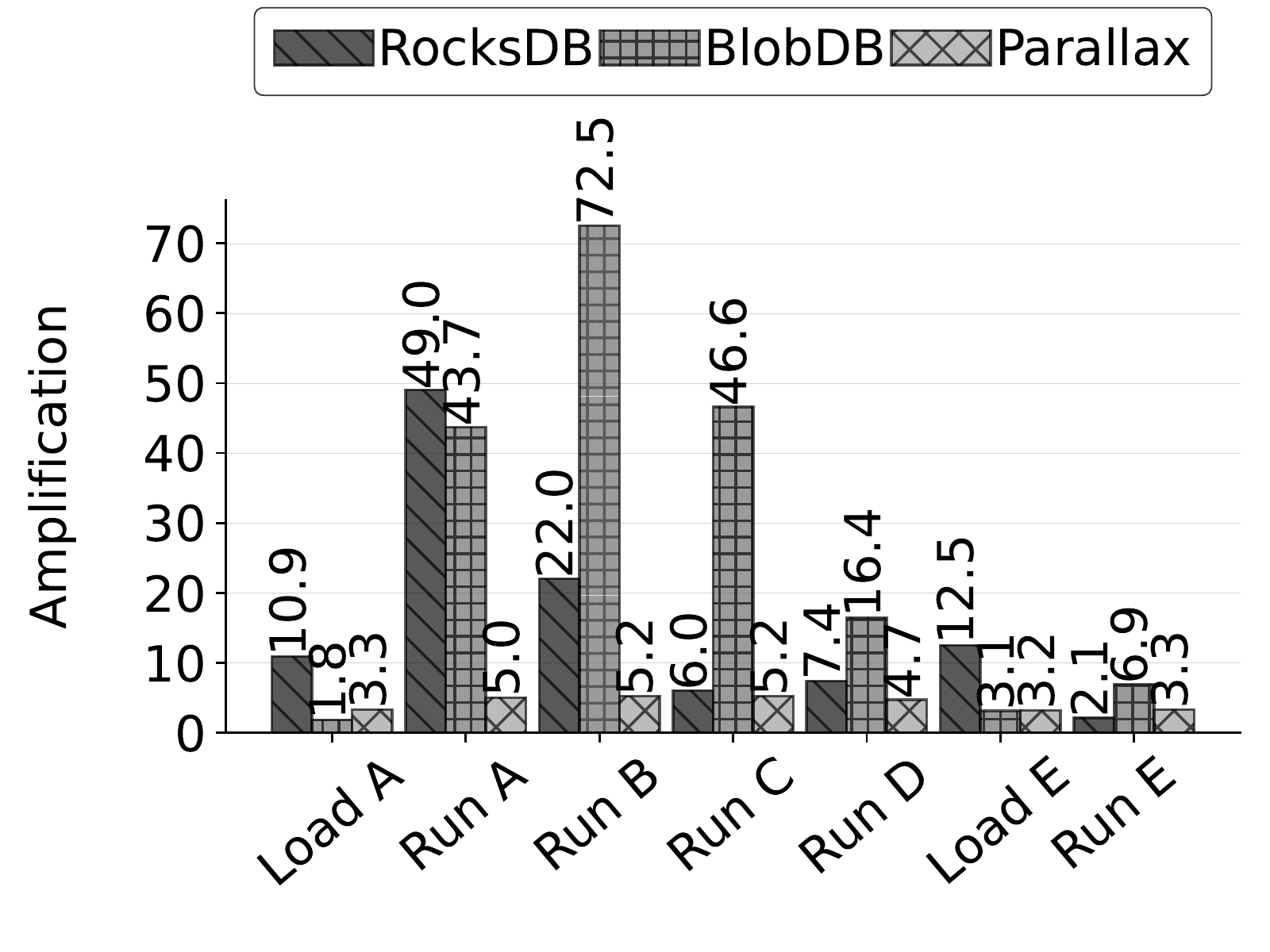}}
\subfloat[Efficiency]{\includegraphics[width=.34\textwidth]{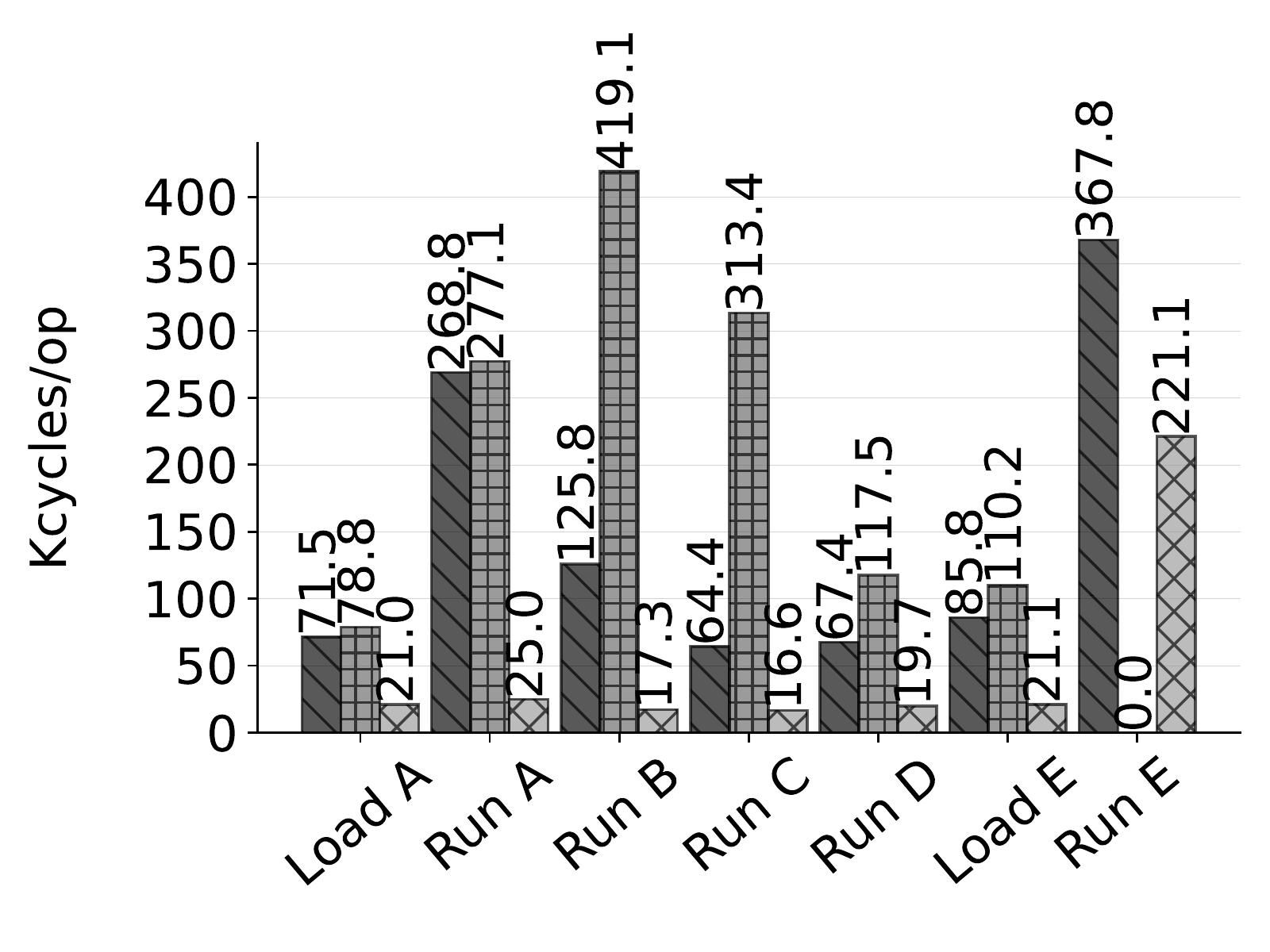}}
\caption{Throughput (left), I/O amplification (middle), and efficiency
  (right) for all YCSB workloads for \name{}, RocksDB, and BlobDB for
  two workloads: SmallD (top row) and MediumD (bottom
  row).}
\label{fig:ycsb}
\end{figure*}

\begin{figure*}[t]
\centering
\subfloat[Throughput]{\includegraphics[width=.34\textwidth]{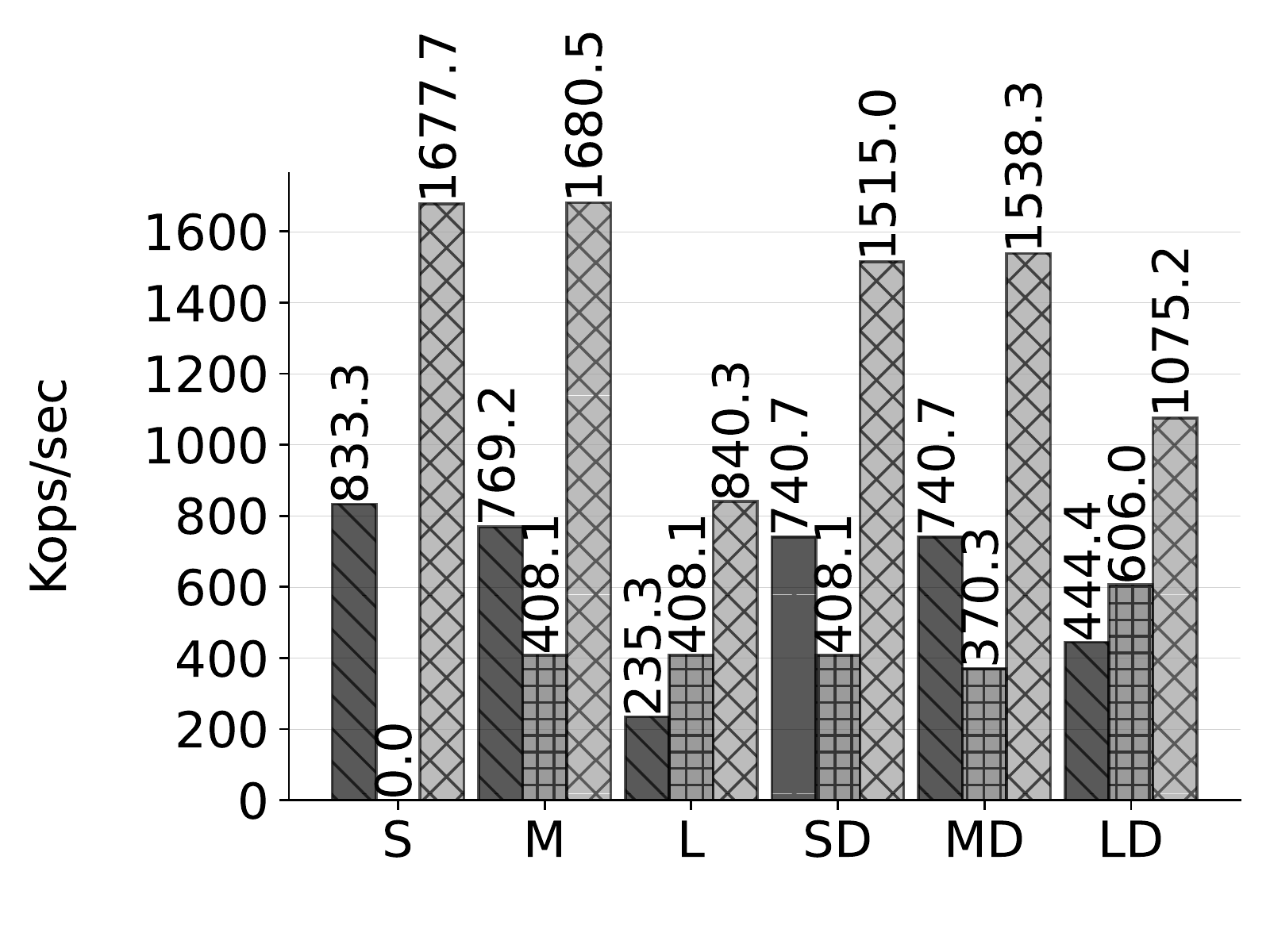}}
\subfloat[I/O Amplification]{\includegraphics[width=.34\textwidth]{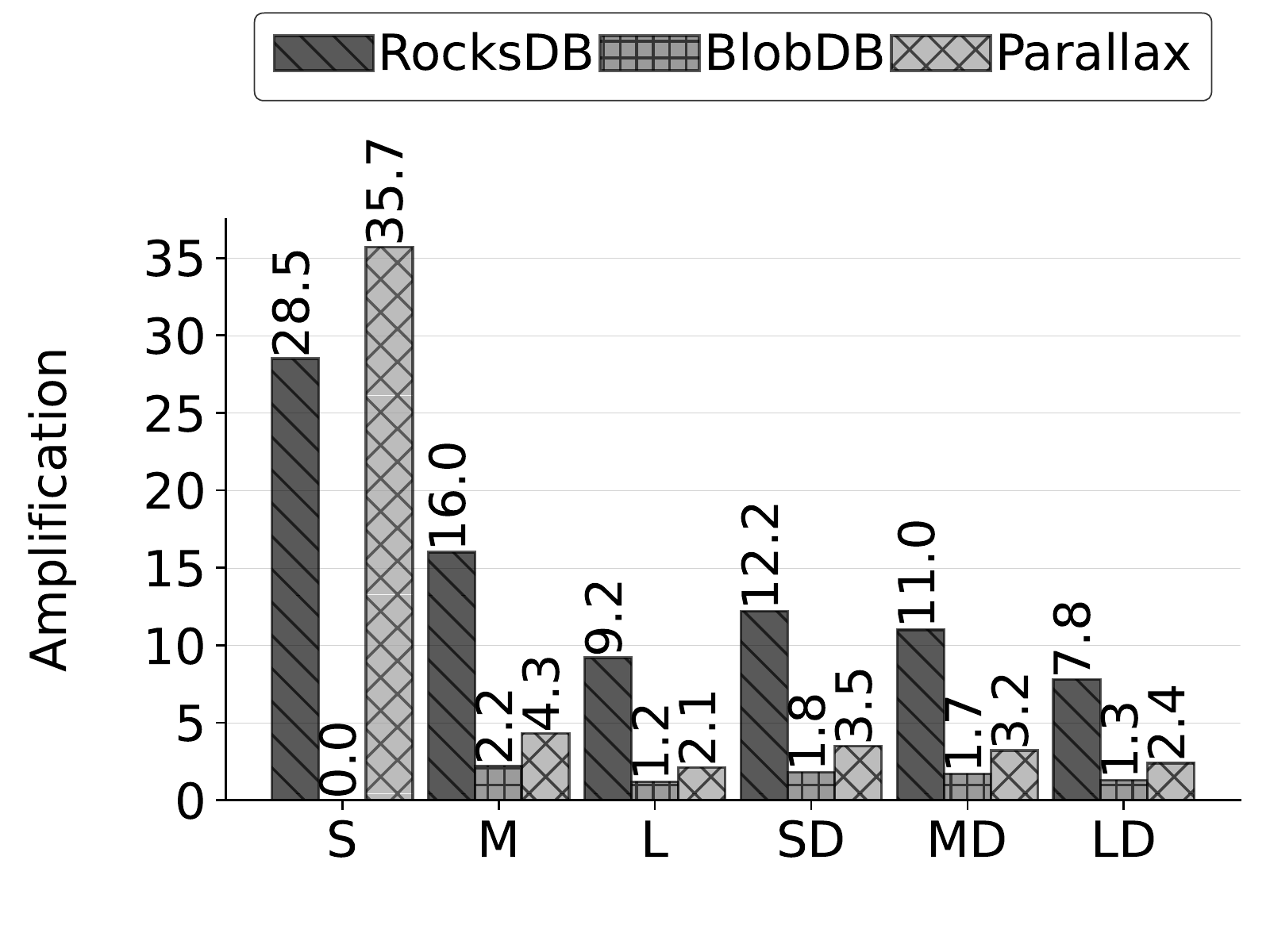}}
\subfloat[Efficiency]{\includegraphics[width=.34\textwidth]{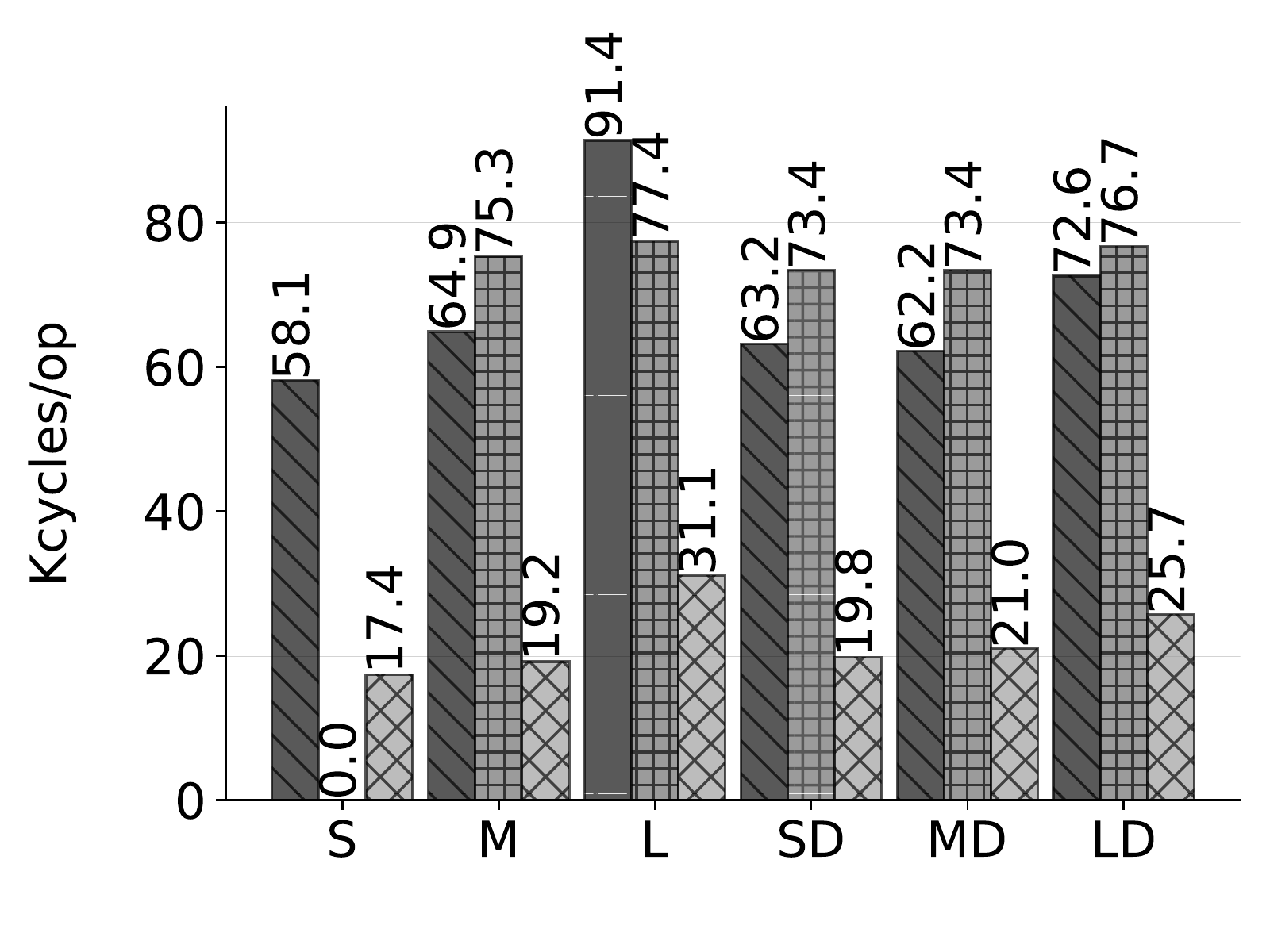}}
\\
\subfloat[Throughput]{\includegraphics[width=.34\textwidth]{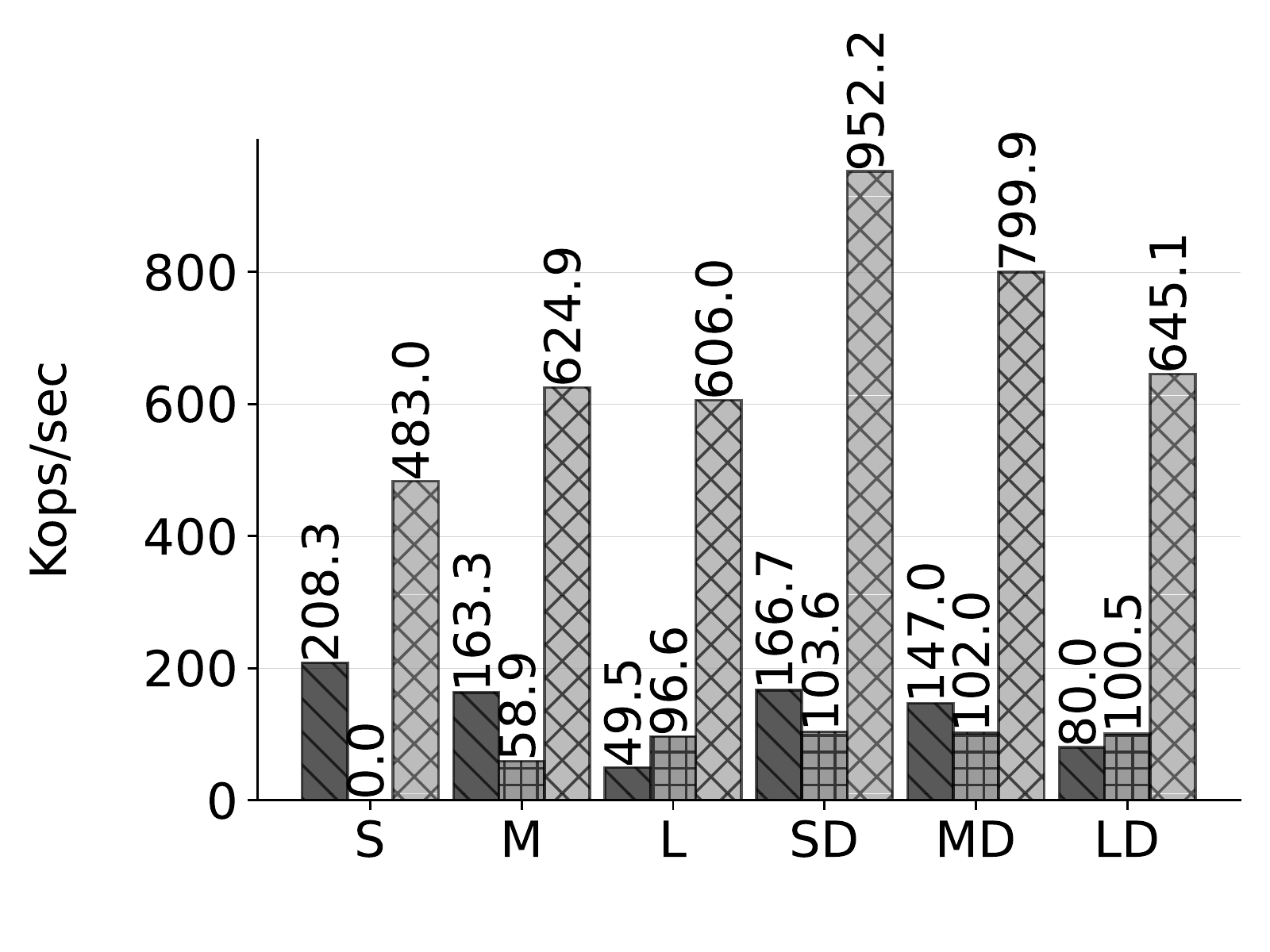}}
\subfloat[I/O Amplification]{\includegraphics[width=.34\textwidth]{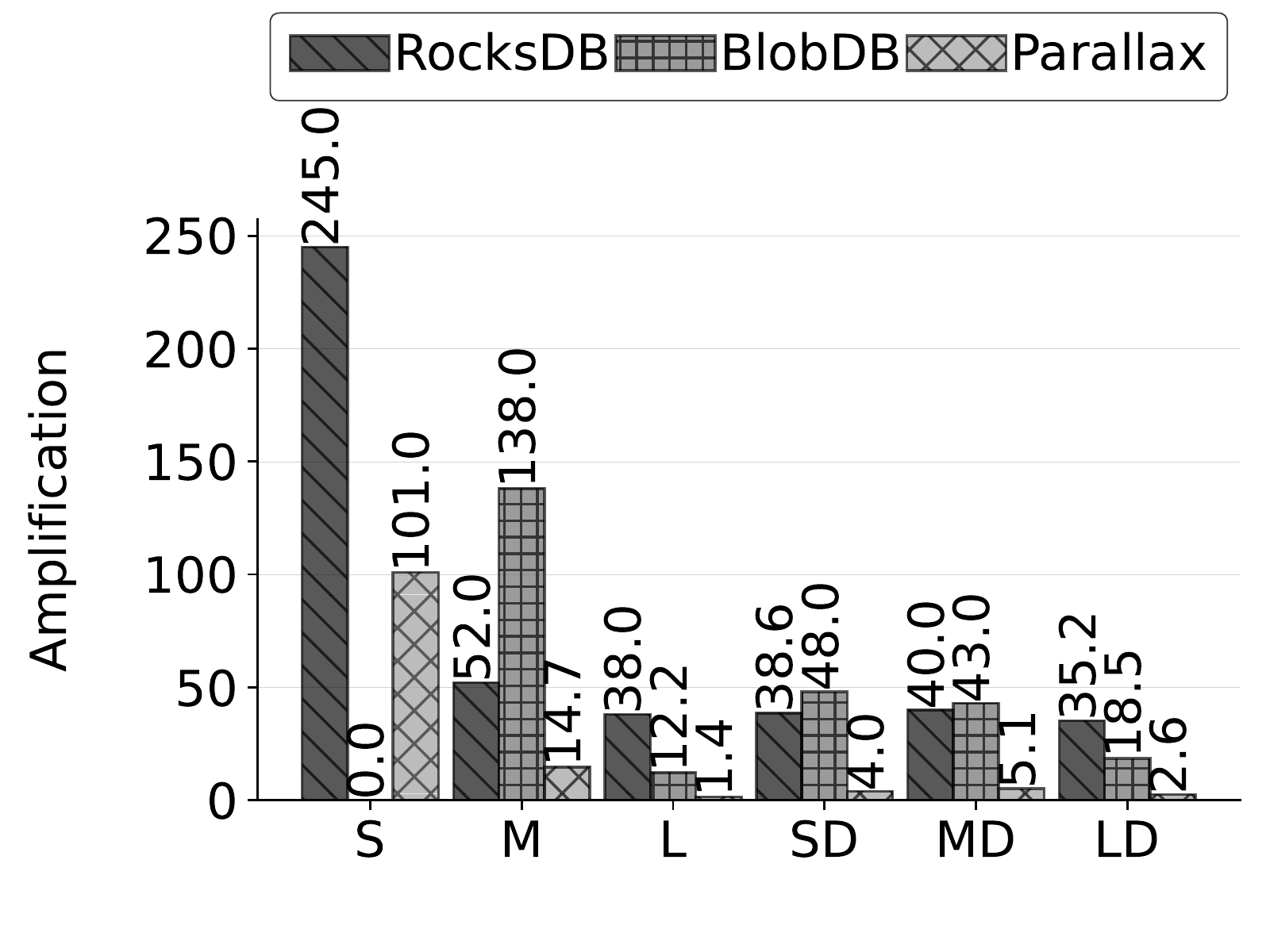}}
\subfloat[Efficiency]{\includegraphics[width=.34\textwidth]{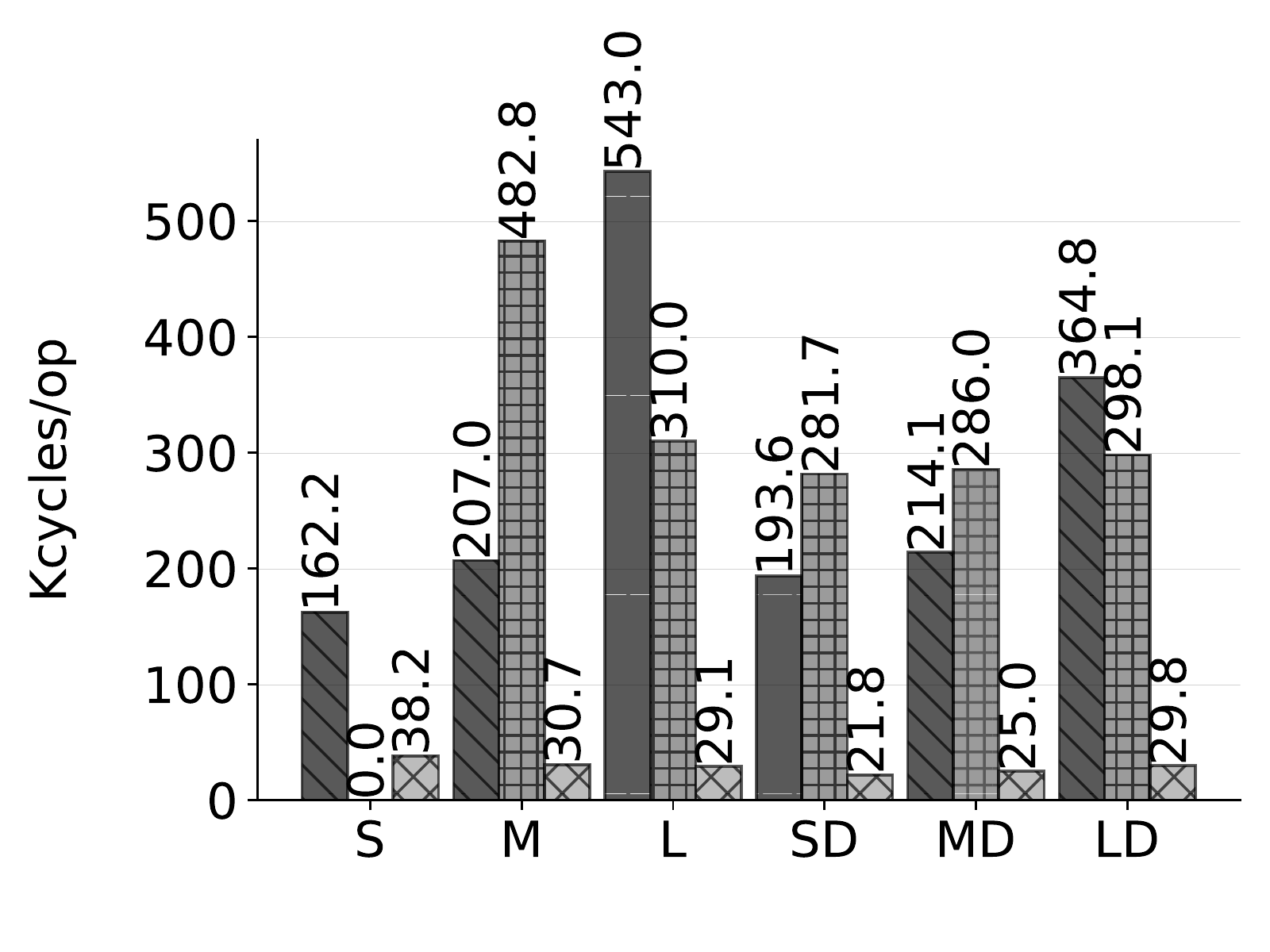}}
\caption{Throughput (left), I/O amplification (middle), and efficiency
  (right) for \name{} (hybrid) compared to RocksDB (always-in-place)
  and BlobDB (always-in-log) for two workloads: Load A (top row) and
  Run A (bottom row).
}
\label{fig:overall}
\end{figure*}

\section{Experimental Evaluation}
\label{sec:eval}

In our evaluation we examine the following questions:

\begin{enumerate}[noitemsep]
\item What is the impact of hybrid KV placement in \name{},
  compared to full in-place and full KV separation?
\item What is the benefit of introducing the medium category? 
\item What is the impact of merging medium KV pairs in-place earlier
  than the last level? 
\item What is the impact of sorting log segments for medium KV pairs
  in $L_0$? 
\end{enumerate}

\paragraph{Impact of Hybrid KV Placement:}
First, Figure~\ref{fig:ycsb} shows results for all YCSB workloads for
SD (top row) and MD (bottom row). We use these two workloads because
they are more typical in modern applications~\cite{258935,246158}.
In addition to Load A and Run A, \name{} increases
throughput, decreases I/O amplification, and increases efficiency for
all workloads except for Run E, compared to both RocksDB and BlobDB. In
both SD and MD, for Run B, C, and D \name{} increases throughput by up
to 5.6x and 61x, reduces I/O amplification by up to 6.5x and 26x, and
increases efficiency by up to 8.6x and 28x.

Run E consists mostly of scan(95\% scans 5\% inserts) operations and we discuss it separately.
 We run Load E and Run E both for SD and MD workloads.
 We run Load E with 100M operations and for  Run E we run 20M operations. In SD, MD 
\name{} moves at least 50\% the medium KVs  from logs to in-place in the index.
Please note, we do not report in the Figures~\ref{fig:overall} efficiency for BlobDB as
this is too high (3165 Kcycles/op) and about 8x worse than both
\name{} and RocksDB. 
Comparing \name{} to RocksDB, RocksDB exhibits 1.43x (SD) and 1.48x
(MD) higher throughput. It is important to note here that having all
keys in place is the best organization for scan operations, which
however comes at a high cost (I/O amplification and CPU efficiency)
for most other workloads. BlobDB has a throughput for SD, MD of 6.2 and 7.6 Kops/s which
makes KV separation impractical for such workloads. However, \name{}
reduces this gap dramatically and within 36\% (SD) and 39\% (MD) of
in-place. Therefore, \name{} hybrid KV placement is a much more
practical approach for KV separation, with significant benefits from
most workloads and small deficiencies for workloads where in-place KVs
perform best.

Next, Figure~\ref{fig:overall} shows in more detail the impact of
hybrid KV placement compared to RocksDB and BlobDB using Load A and
Run A for all six mixes of KV-pair sizes (Table~\ref{tab:workloads}).
Note, that the GC cost for \name{} in Load A is almost zero due to its GC mechanism (no updates take place), while \name{} exhibits the full GC cost for Run A.

For Load A (Figure~\ref{fig:overall}, top row), we see that in terms
of operation throughput, \name{} exhibits up to 3.57x and 4.15x higher
throughput compared to RocksDB and BlobDB, respectively. For all
workloads except S (small KV pairs), \name{} reduces I/O amplification
by up to 4.38x compared to RocksDB. Compared to BlobDB, \name{} exhibits
up to 1.95x higher I/O amplification because BlobDB places all values in
a log and never includes values in compactions.  However, GC cost in
BlobDB is high and makes BlobDB throughput significantly worse
compared to \name{}. 
In terms of CPU efficiency we observe that \name{} is
by up to 3.92x better than both systems, with RocksDB and BlobDB
generally being relatively close and within 17\% of each other.
Please note that in the case of large only \name{} exhibits slightly
higher I/O amplification compared to BlobDB (2.1 vs 1.2) due to its
B+-tree index per level.
For workload S, \name{} has 1.25x worse amplification because of the
slot array in the B+-tree leaves.  In particular, when KVs are small,
the slot array accounts for 8\% of the total leaf's
capacity. Regarding throughput \name{} has 2x more throughput than
RocksDB. We suspect this difference comes from 1) The difference of
the in-memory component~\cite{jellyfish,flodb} and 2) RocksDB has
resilience mechanisms such as CRCs (which cannot be disabled). We
leave this investigation for future work. However, in our evaluation,
we focus mostly on amplification which mainly represents \name{}
techniques.

For Run A (Figure~\ref{fig:overall}, bottom row) \name{} improves
throughput, I/O amplification, and efficiency even further.
Run A exhibits both lookup and cleanup costs for
GC in the log for BlobDB, therefore, I/O amplification in
BlobDB becomes even worse compared to \name{}. \name{} compared to RocksDB and
BlobDB, increases throughput by up to 12.24x and 10.75x, reduces I/O
amplification by up to 27.1x and 9.38x, and increases efficiency by up to
18.7x and 16x. For Run A \name{} has 1.3x less throughput and 2.4x more amplification.

\begin{figure*}[t]
\centering
\subfloat[Throughput]{\includegraphics[width=.34\textwidth]{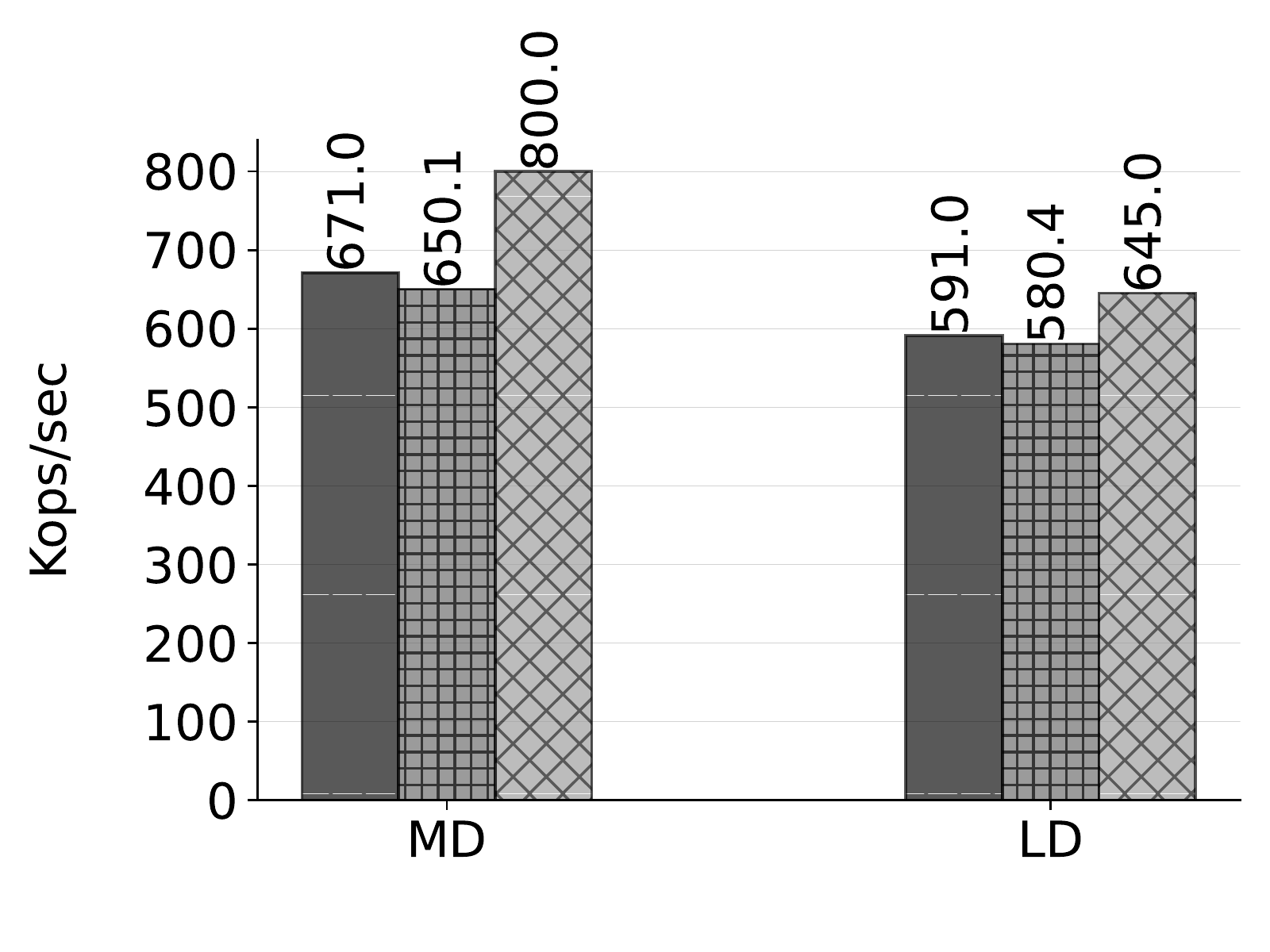}}
\subfloat[I/O Amplification]{\includegraphics[width=.34\textwidth]{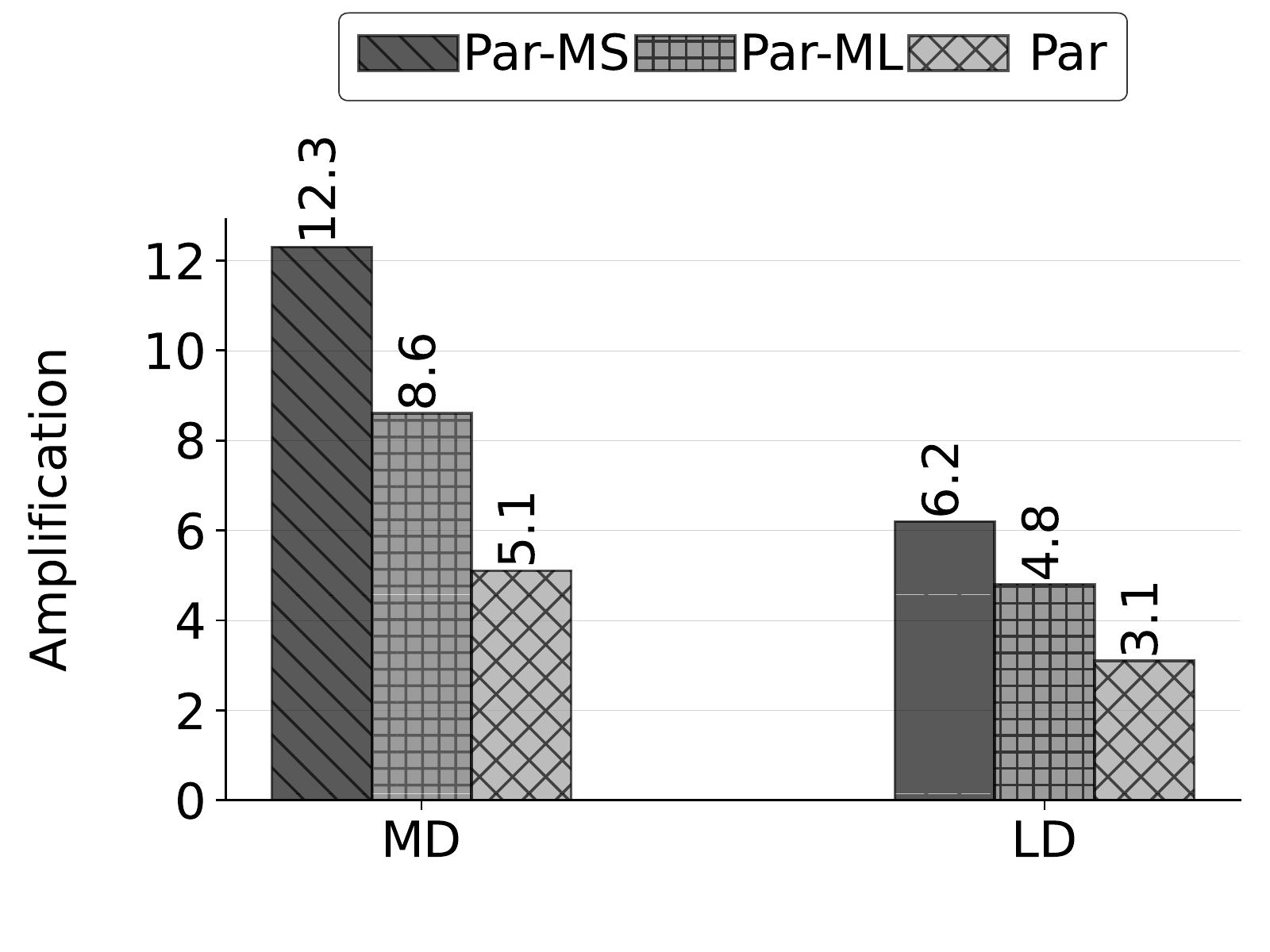}}
\subfloat[Efficiency]{\includegraphics[width=.34\textwidth]{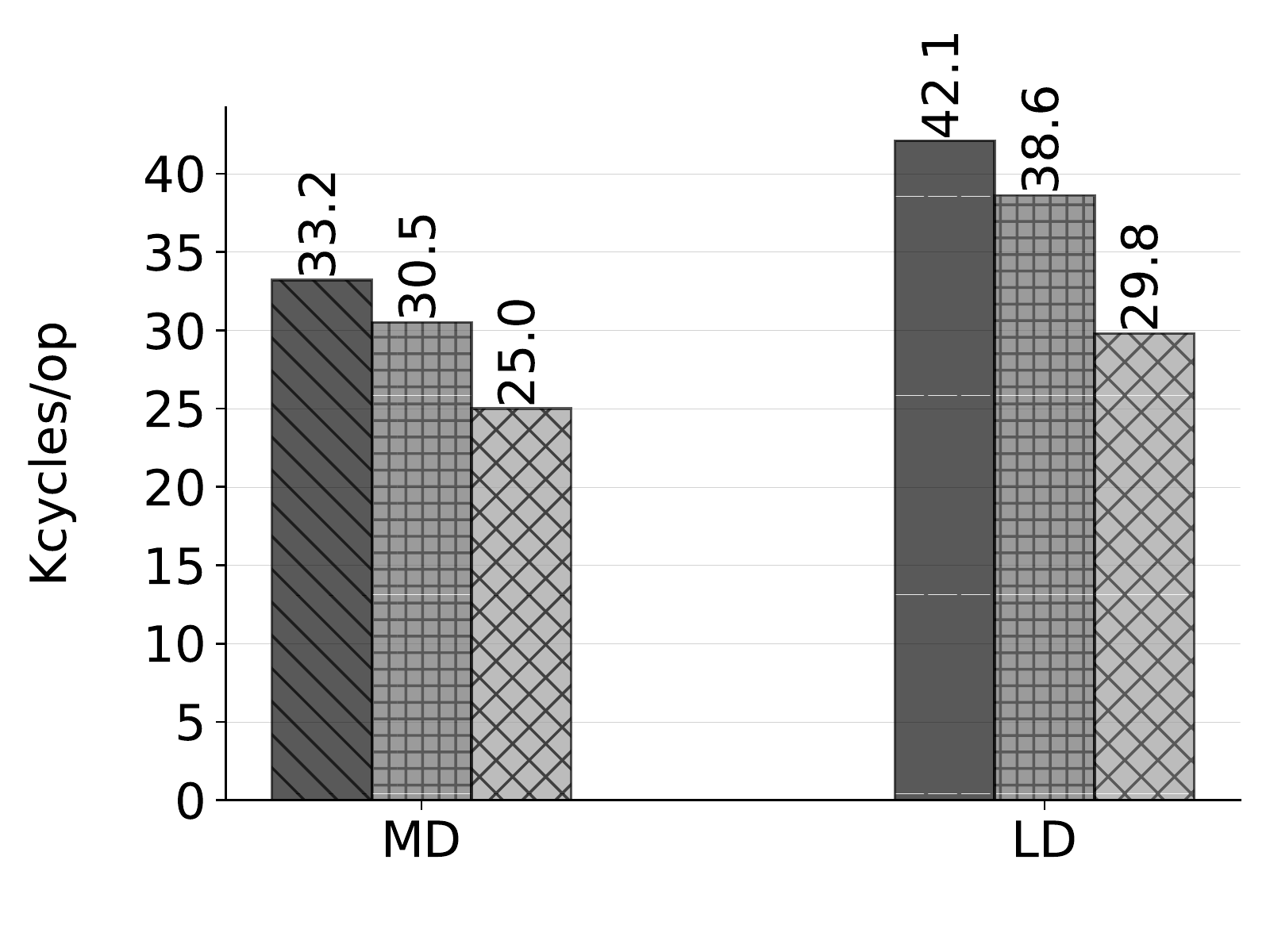}}
\caption{Run A throughput, I/O amplification, and efficiency for
  three configurations:  1) \name{}-MS and 2)\name{}-ML and 3) \name{} (Par). \name{}-MS and \name{}-ML use only the small and large
  categories.  \name{}-MS classifies medium keys as small, whereas
  \name{}-ML classifies them as large.}
\label{fig:cats2vs3}
\end{figure*}

\paragraph{Benefits of introducing the medium category:}

Next, we examine if the complexity of introducing the medium category
results in substantial benefits.  Figure~\ref{fig:cats2vs3} shows
throughput and I/O amplification for Run A for two \name{}
configurations: moving medium KV pairs to the small category
(\name{}-MS) and moving them to the large category (\name{}-ML).
Essentially, these two configurations for \name{} correspond to
setting the KV pair thresholds to $T_{SM} = T_{ML} = 0.02,$ for
\name{}-MS and to $T_{SM} = T_{ML} = 0.2$ for \name{}-ML.

We configure GC as we describe in Section~\ref{sec:method}. 
Also, the duration of the experiment is such that in all cases
the GC threads process at least 97\% of the system logs by the time
the workload finishes. Figure~\ref{fig:cats2vs3} shows that for MD and
LD using \name{} compared to \name{}-MS and \name{}-ML improves
throughput by up to 1.23x and 1.11x respectively. Also, it reduces I/O
amplification by up to 2.43x and 2x and increases efficiency by up to 1.32x
and 1.41x.  As expected, the difference is higher for MD, since in LD
the large percentage of large KV pairs results in all three \name{}
configurations placing most of the dataset in the log for large KV
pairs.

\begin{figure*}[t]
\centering
\subfloat[Throughput]{\includegraphics[width=.34\textwidth]{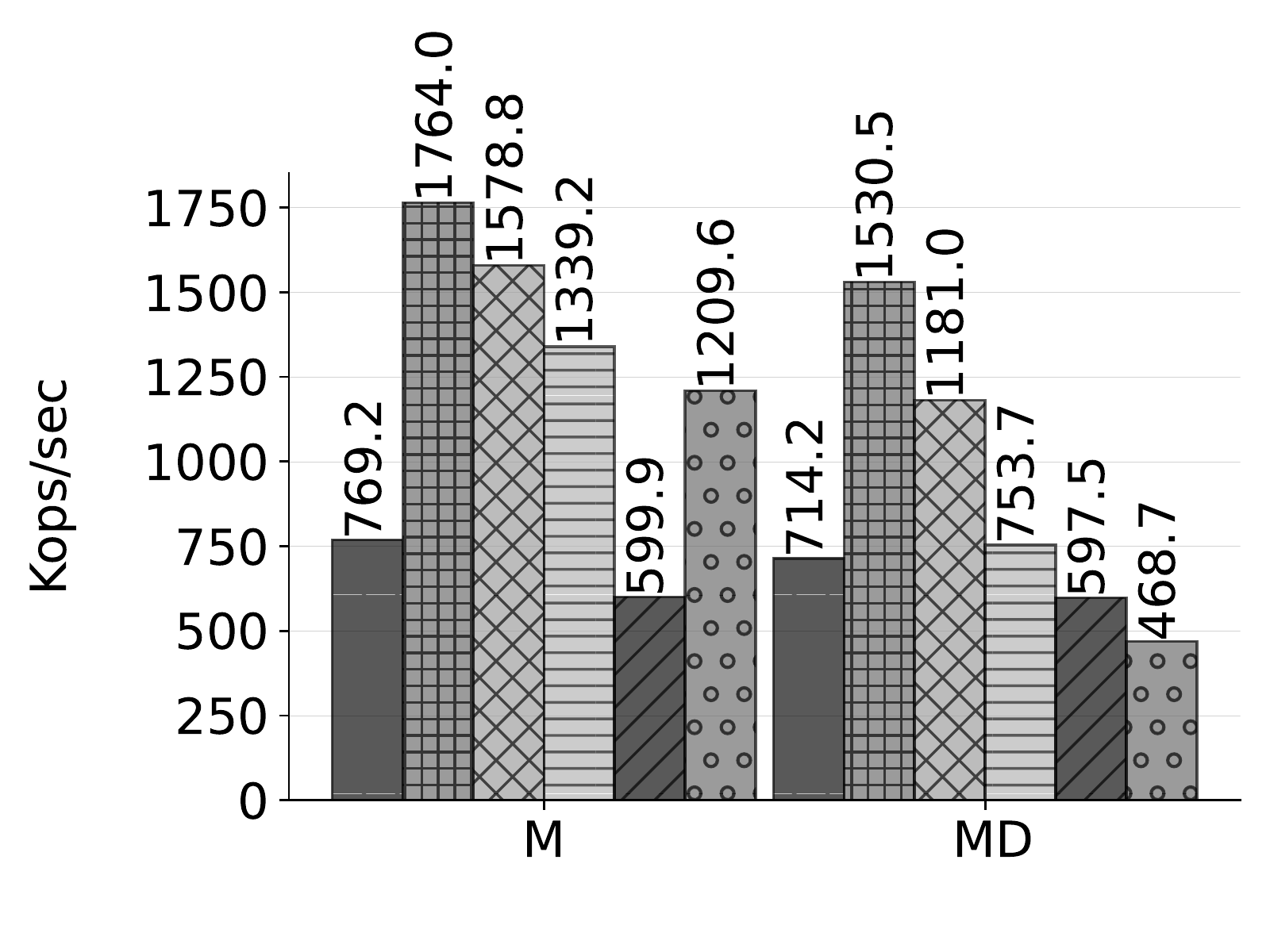}}
\subfloat[I/O Amplification]{\includegraphics[width=.34\textwidth]{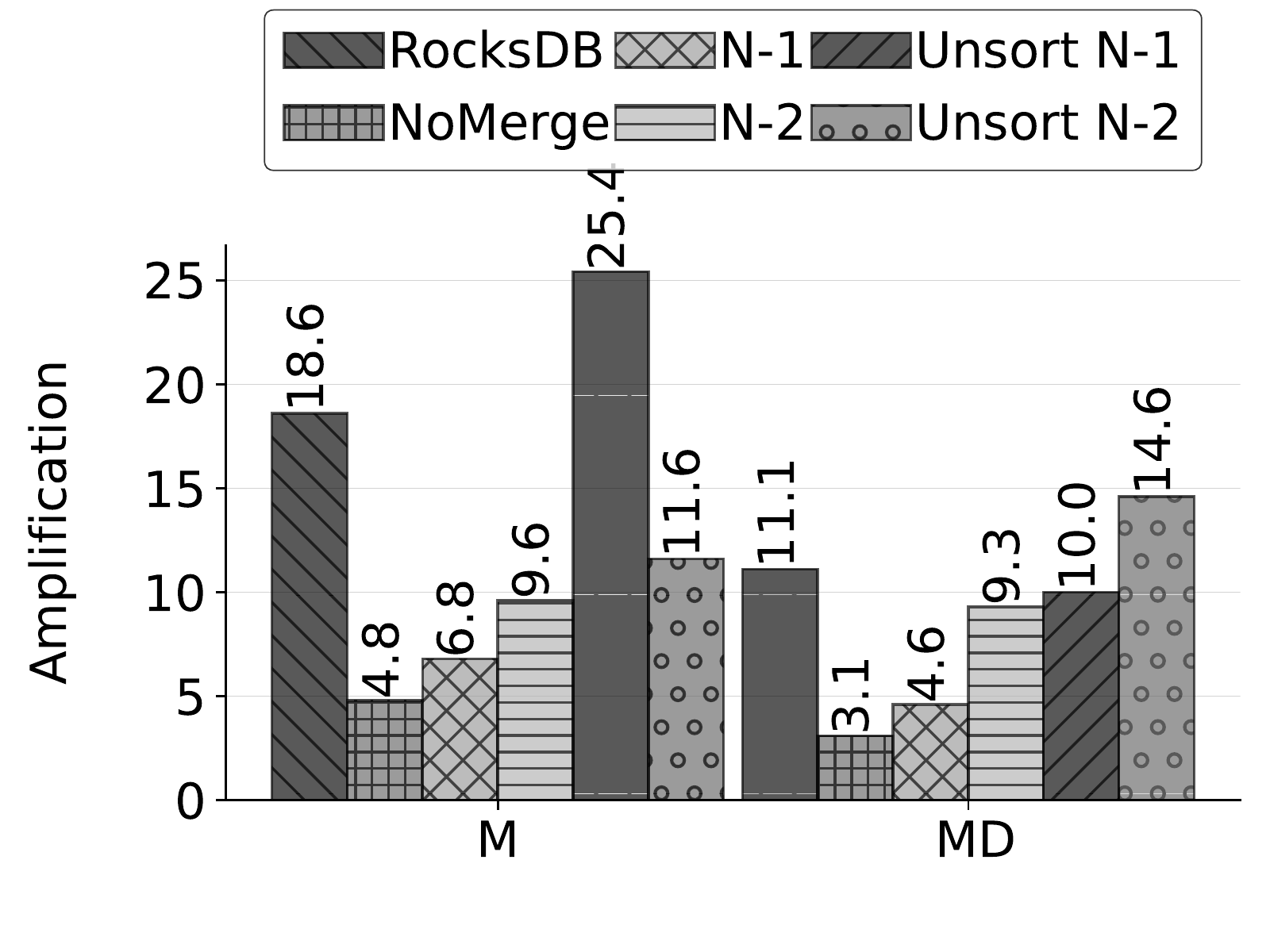}}
\subfloat[Efficiency]{\includegraphics[width=.34\textwidth]{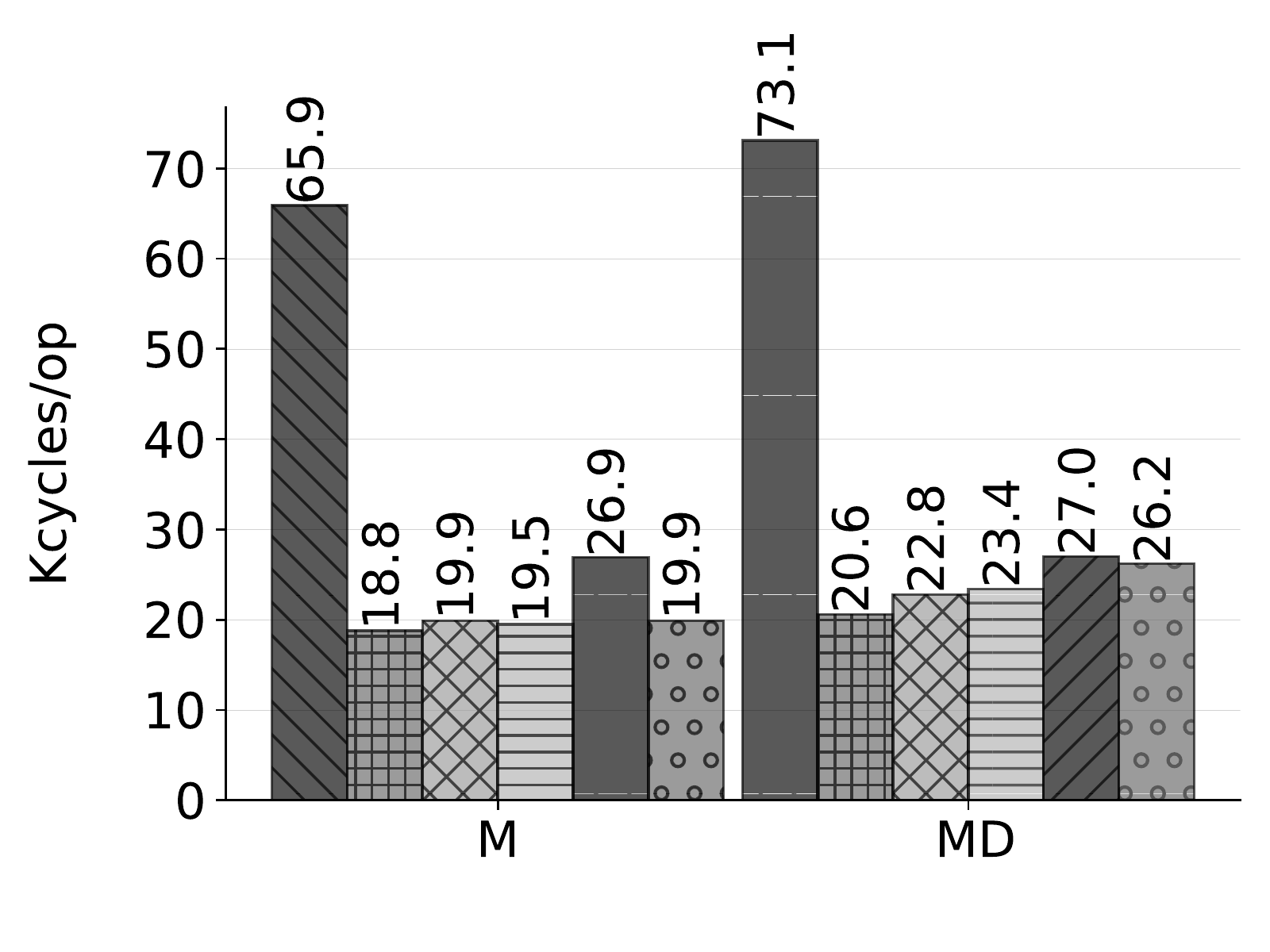}}
\caption{Impact of merging medium KV pairs earlier for Load A that exhibits more compactions.(Top Row)
Impact of sorted $L_0$ segments for Load A that exhibits more compactions. (Bottom Row)}
\label{fig:mergelevel}
\end{figure*}

\paragraph{Merging medium KV pairs in-place earlier:}

\name{} can merge medium KV pairs in place at different levels of the
LSM structure. Merging KV pairs later in the LSM structure results in
lower I/O amplification but higher space amplification due to larger
logs for medium KV pairs.  Therefore, merging medium KV pairs earlier
limits the size of the transient log and thus reduces space amplification.

In terms of space amplification, our model
(Figure~\ref{fig:valuelog}(b)) shows that merging medium values
in-place in $L_{n-2}$ vs $L_{n-1}$, reduces space amplification for a
growth factor of 8 from about 13\% to less than 3\%. 
For a growth factor of 4, space amplification is reduced
from 25\% to 6\%. In terms of I/O amplification, we observe that each
level contributes equally to level amplification (one level less for
compactions of medium KV pairs), therefore, moving medium values
in-place one level earlier increases level amplification by 1/N (N is
the max number of levels based on storage capacity). However, total
I/O amplification includes also merge amplification.

In Figure~\ref{fig:mergelevel} we configure Parallax with growth factor 4 and set $L_0$
size to 128MB. We run Load A with 150M keys with workload M to create enough levels in the LSM tree and
stress the system. In the runs with labels (N-1)/(N-2), Unsort (N-1)/(N-2), \name{} has transferred at least 90\% of medium keys in place.
Figure~\ref{fig:mergelevel} quantifies the impact of merging medium KV
pairs earlier. If we examine merging medium
KV pairs (M workload - (N-1)/(N-2)) at levels $L_{N-1}$ vs $L_{N-2}$, I/O amplification is 6.8
vs. 9.6 and throughput is 1579 Kops/s vs. 1339 Kops/s
(Figure~\ref{fig:mergelevel}(a,b)).  Therefore, a 16\% improvement in
throughput and 34\% improvement in I/O amplification come at a cost of
about 4x increase in space amplification (from 6\% to 25\% for growth
factor 4 or from 13\% to 3\% for growth factor 8).  Depending on the
tradeoffs in specific setups, we believe that merging values at
either of the last two levels can be a good approach. Especially, if
scan performance is also important, then merging at $L_{n-2}$ is a
good tradeoff.

As a reference point, we also include numbers for RocksDB and a
non-achievable (ideal) baseline, NoMerge. NoMerge is a version of
\name{} that keeps medium KV pairs always in the log and never merges
them in place, however, without performing any GC in the log either.

\mycomment{

\begin{figure}[t]
\centering
\includegraphics[width=.90\columnwidth]{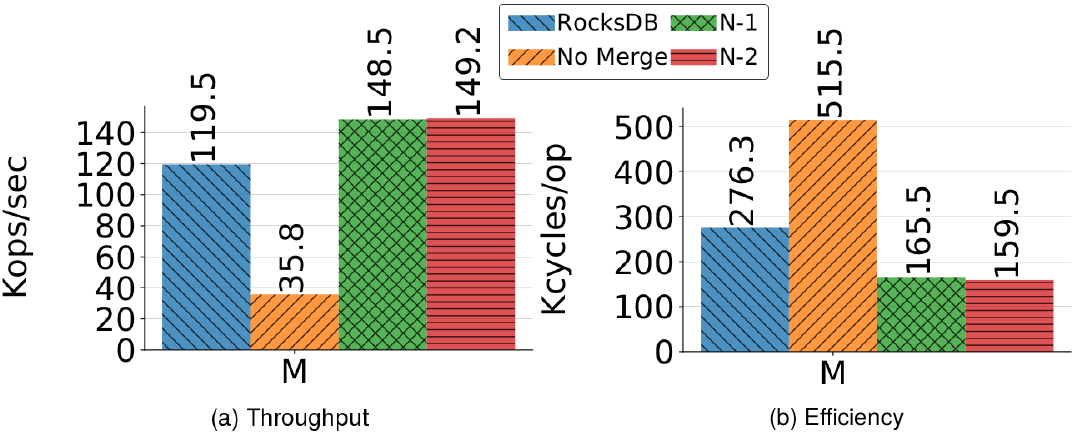}
\caption{Impact of having medium keys in log or in place for Run E.}
\label{fig:medscansinplace}
\end{figure}

}

\paragraph{Impact of sorting log segments in $L_0$:}
To reduce the I/O traffic generated when merging medium KV pairs in
place, \name{} uses a technique that eagerly sorts each transient log
segment (as shown in Figure~\ref{fig:medium_log_compaction}).  This
technique ensures that \name{} fetches only once each transient log
segment in memory before merging in-place.

Figure~\ref{fig:mergelevel} shows that in workload M, sorting segments
in $L_0$ (when merging medium KV pairs at $L_{N-1}$), improves
throughput by up to 2.63x, from 600 to 1578 Kops/s, and reduces
I/O amplification by up to 4x (from 25.8 to 6.8). As a secondary
observation, we note that if we choose to use unsorted segments, it is
preferable to merge medium KV pairs at $L_{N-2}$ instead of $L_{N-1}.$
For workload MD, using sorted segments results also in higher
throughput by up to 1.92x (merging at $L_{N-1}$), in higher efficiency by
up to 1.2x, and
2.17x higher I/O amplification.  Therefore, sorted
transient log segments appear to be overall a better approach. 


\mycomment{
Next, we configure \name{} to use always-in-place, always-in-log, and
hybrid (default) policies.  We see that \note{fill in}.  \note{ab:
  What does this figure tell us in addition to the rocksdb/blobdb
  discussion?}  \note{ab: For small, throughput, why is inlog better
  than small/hybrid? This is the case where the log should be
  bad... Also, in Run A, small/throughput should be really bad for
  inlog but in this figure it is only a bit worse...}

\begin{figure}
\centering
\subfloat[Throughput]{\includegraphics[width=.25\columnwidth]{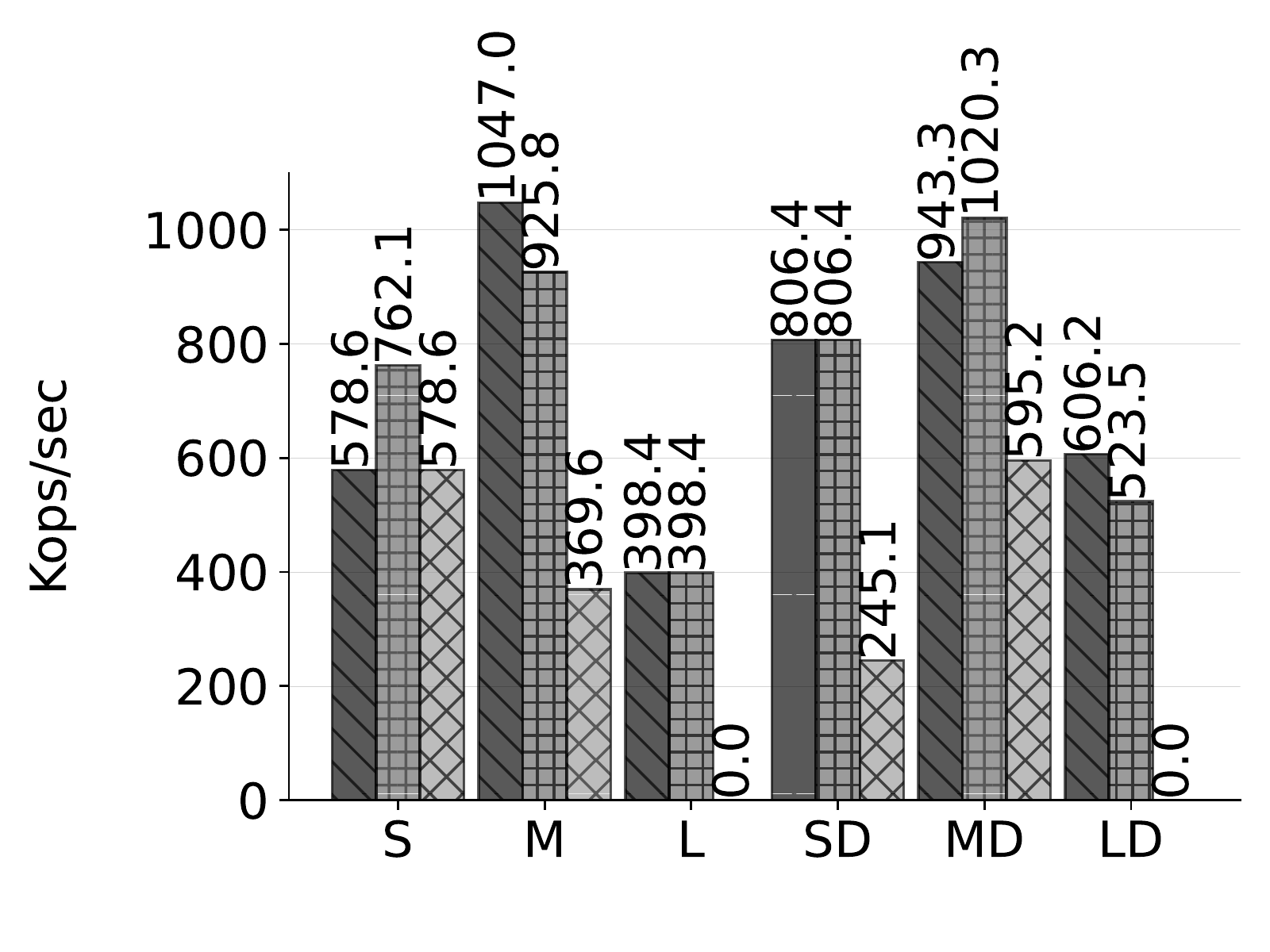}}
\subfloat[I/O Amplification]{\includegraphics[width=.25\columnwidth]{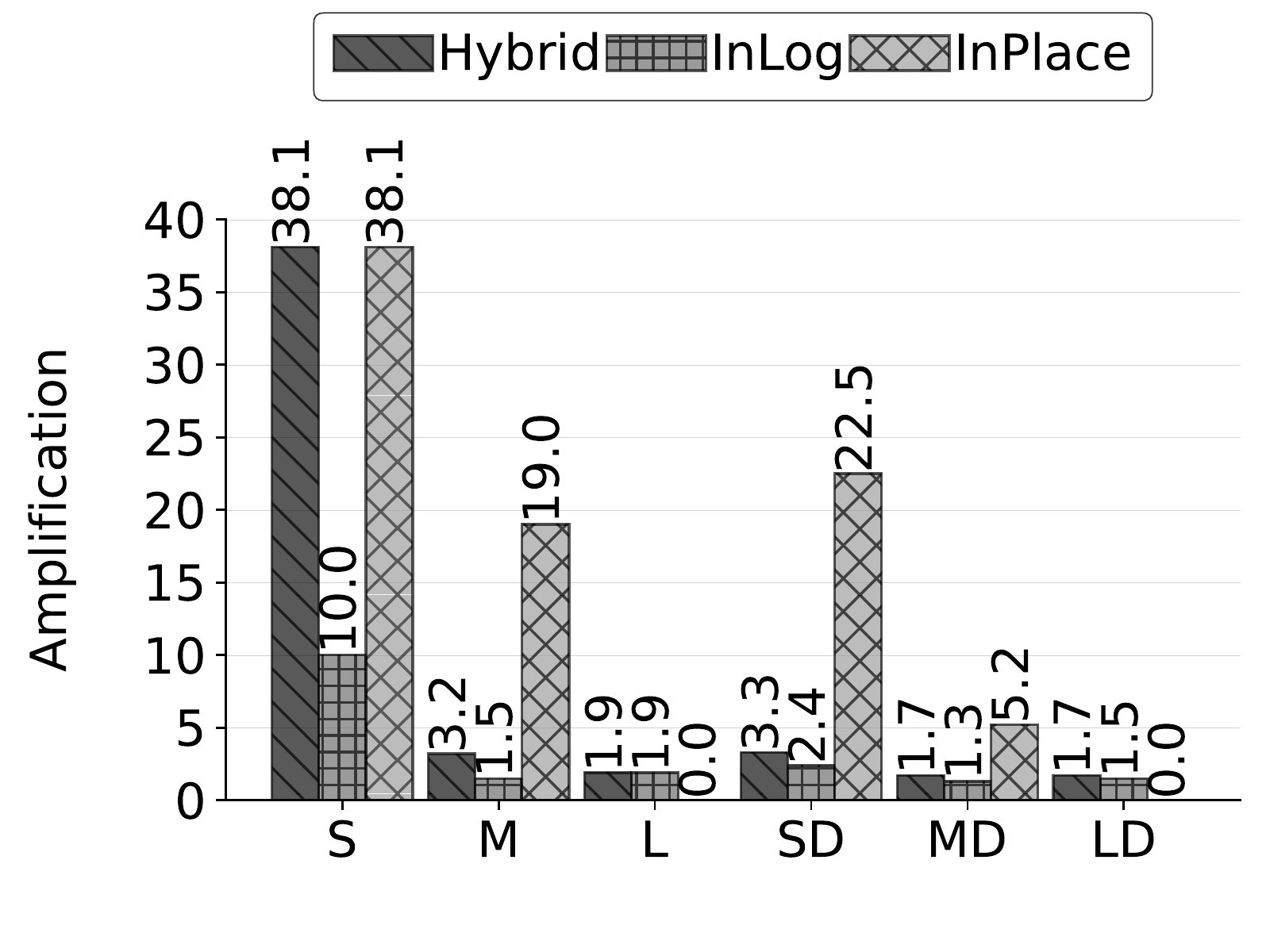}}
\subfloat[Efficiency]{\includegraphics[width=.25\columnwidth]{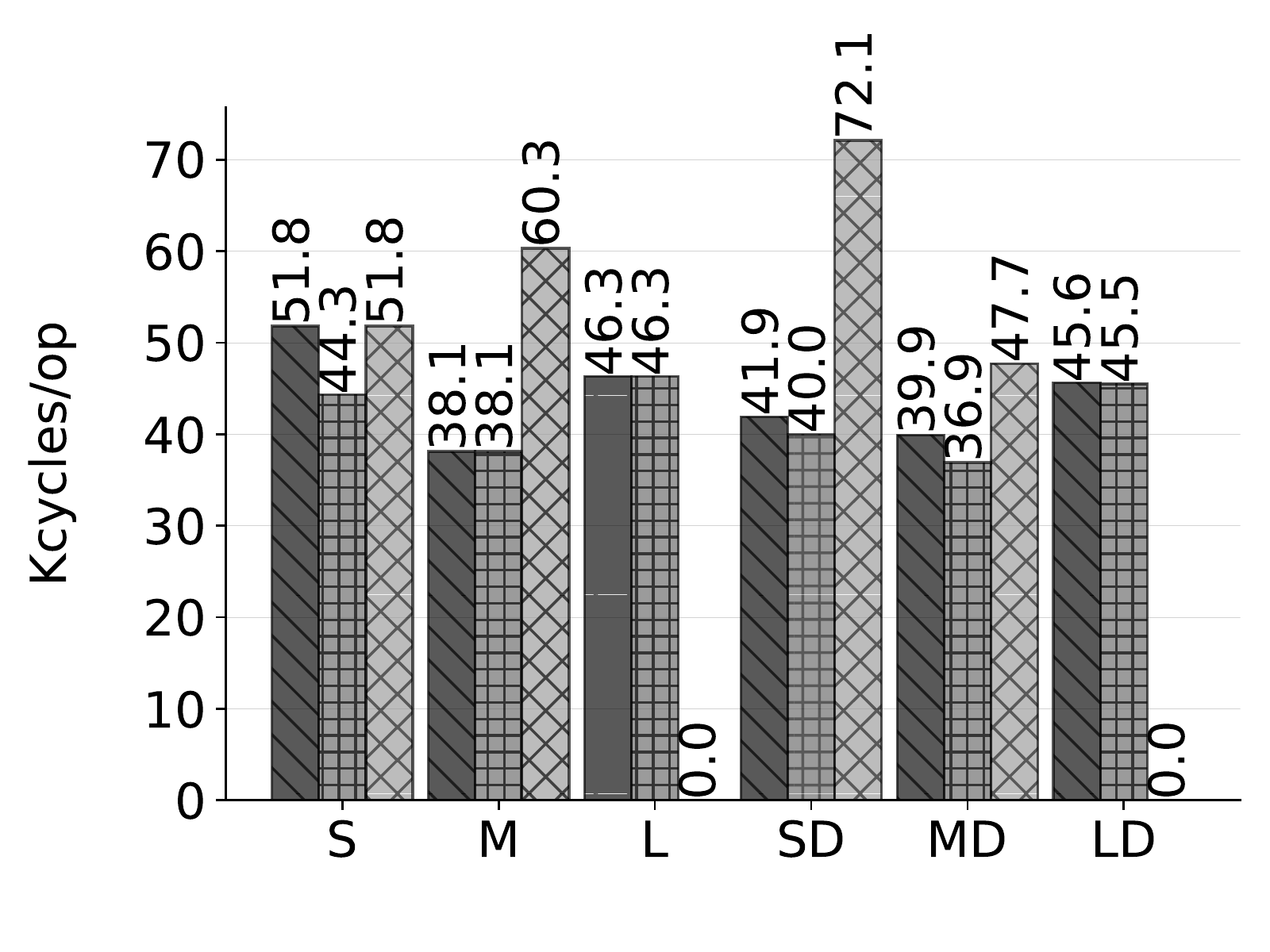}}
\\
\subfloat[Throughput]{\includegraphics[width=.25\columnwidth]{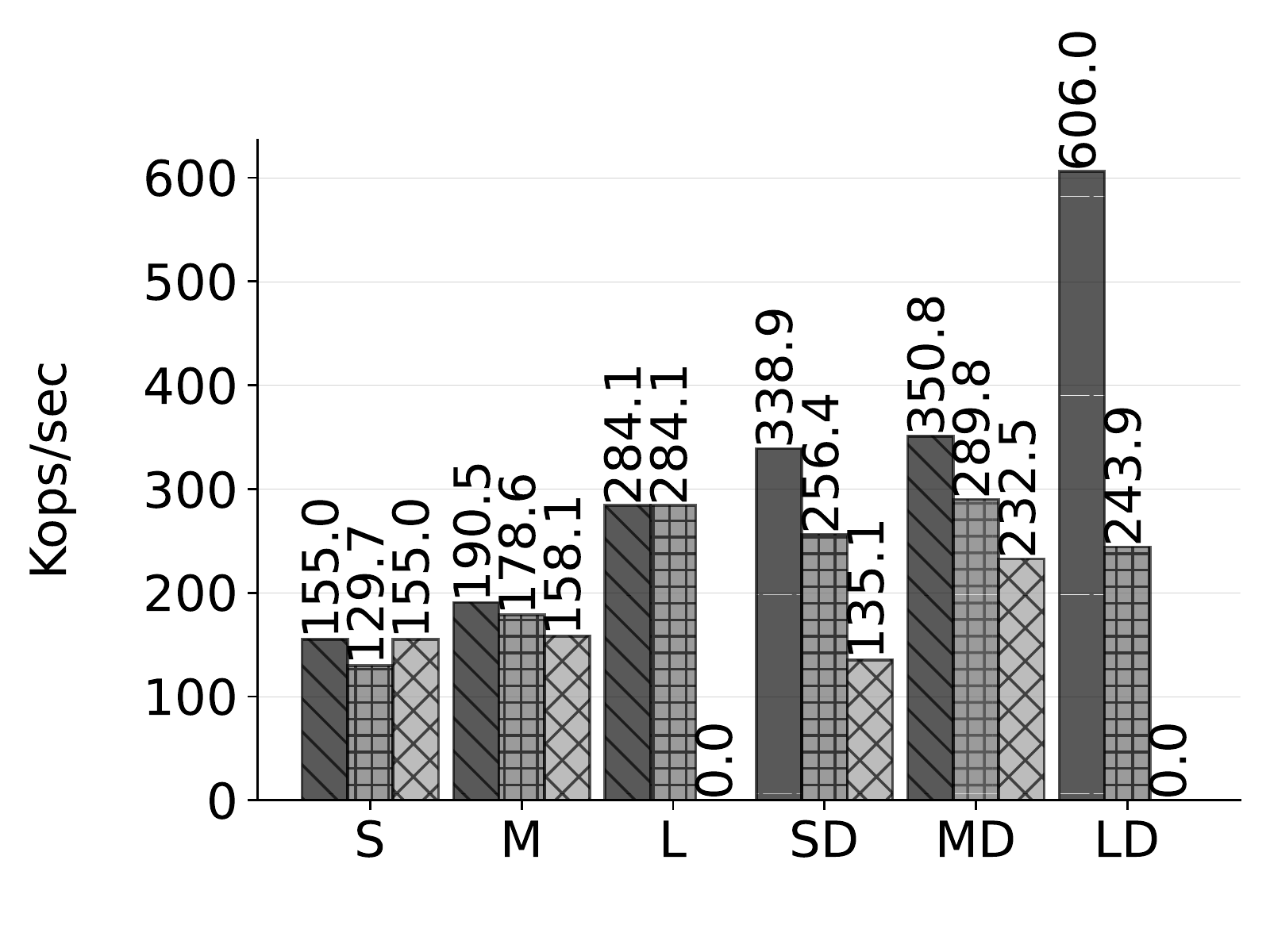}}
\subfloat[Amplification]{\includegraphics[width=.25\columnwidth]{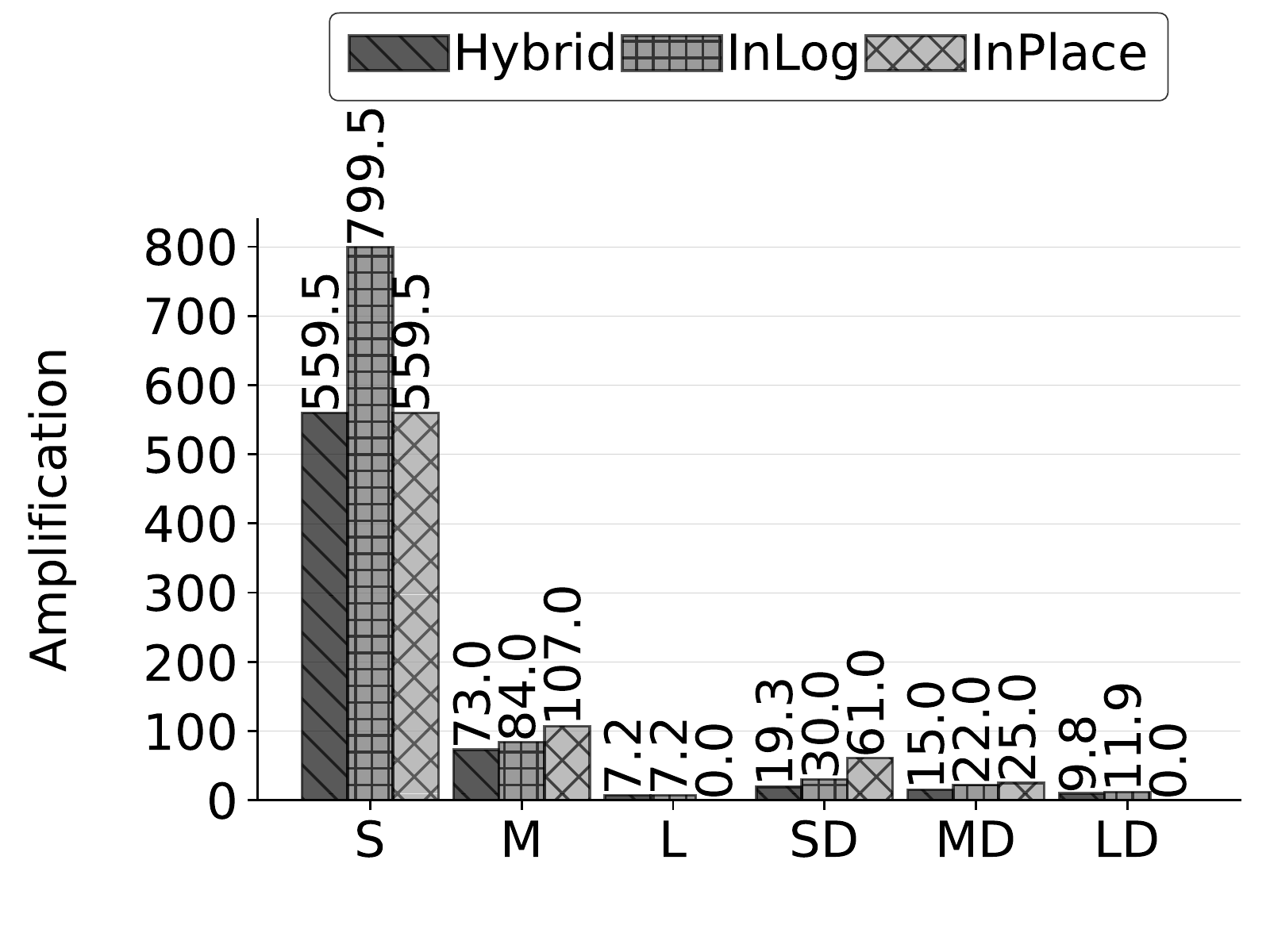}}
\subfloat[Efficiency]{\includegraphics[width=.25\columnwidth]{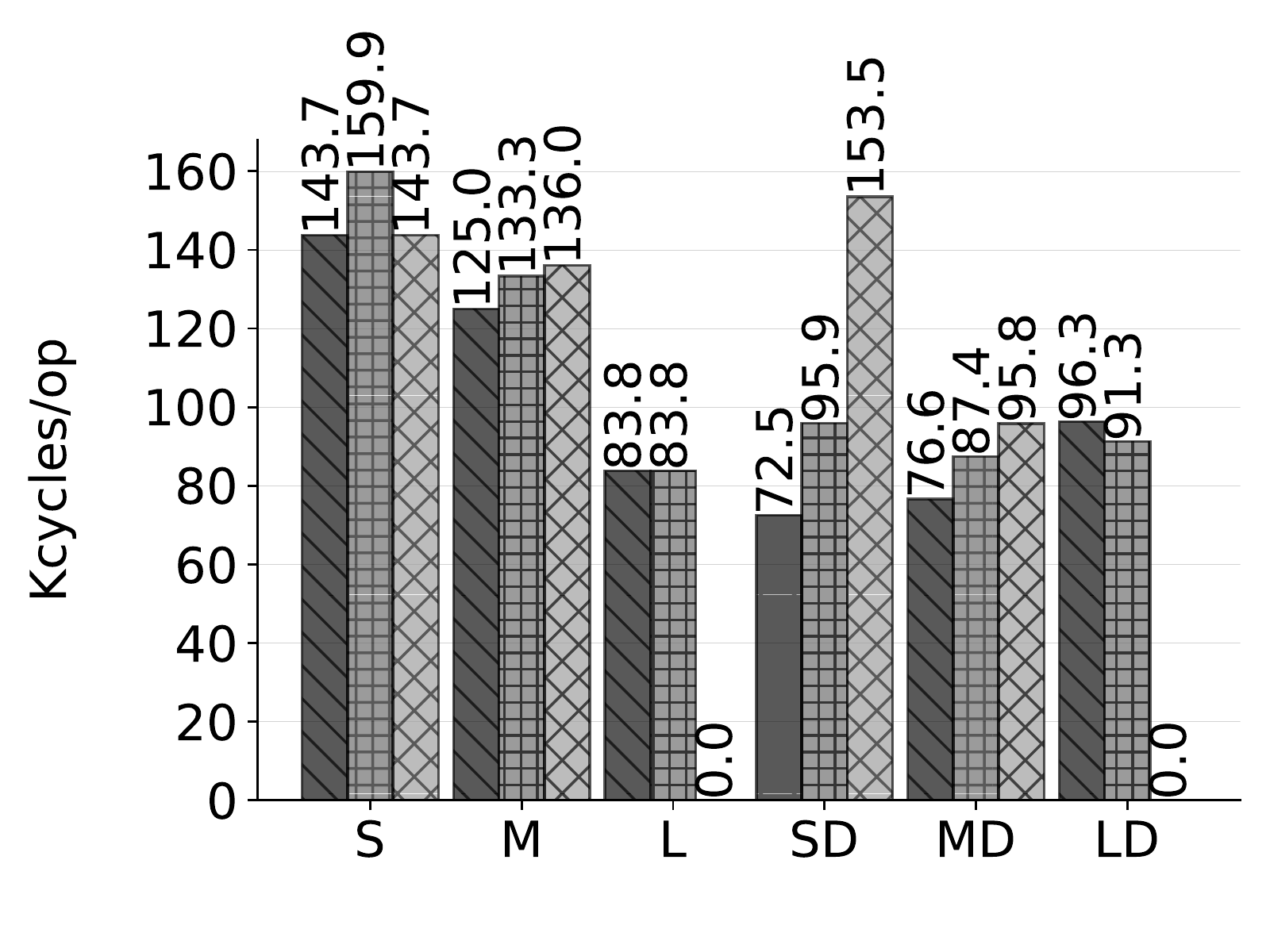}}
\caption{Throughput (left), I/O amplification (middle), and efficiency
  (right) for three configurations of \name{} (hybrid,
  always-in-place, always-in-log) for two workloads: Load A (top row)
  and Run A (bottom row). 
  \note{ab: in load A throughput, why is in-log so good for smallD,
    medD? is it possible that there is pracically no GC taking place
    which favors in-log?}}
\label{fig:parallax}
\end{figure}

}
 
\section{Related Work}
\label{sec:related}

In this section we group related work in the following categories: (a)
Techniques for KV separation, (b) GC for KV separation, and (c) Other
techniques for reducing I/O amplification in LSM-Tree KV stores:

Previous work~\cite{lsmmodel} models write amplification
taking into account update operations.  Our analysis does not consider
updates, but we rather focus on modeling write amplification taking
into account KV separation, which is the basis of Parallax.
Previous systems that employ KV separation, such as
Atlas~\cite{atlas}, BlobDB~\cite{blobdb}, WiscKey~\cite{wisckey},
HashKV~\cite{hashkv} and Kreon~\cite{kreon}, append KV pairs in a
value log and organize their index as a leveled LSM structure. At each
level they keep only the metadata to the actual value
locations. Tucana~\cite{tucana} also performs KV separation and uses a
different multistage index structure of a \betree{}~\cite{betree}. It
stages KV pairs only at the last level of the index to reduce I/O
amplification assuming a certain DRAM/Flash capacity
ratio. KVell~\cite{kvell} is an efficient log structured key-value
store designed for fast storage devices. It uses a value log and a
single level B+-tree index in which it stores metadata (pointers) to
the actual key-value pairs. Furthermore, it uses asynchronous IO
(io\_uring~\cite{Axboe:2019}) and batching to improve efficiency of
device I/O. All these systems suffer from high GC overhead, especially
for small and medium KV pairs which are important in production
workloads~\cite{246158}. Furthermore, the benefits of performing KV
separation for small key value pairs, even without considering the GC
costs, is practically negligible.

HashKV~\cite{hashkv} deals with the high cost of GC in leveled
LSM-type KV stores that perform KV separation.  HashKV reduces GC
overhead for update intensive workloads with heavy zipfian probability
distribution. It tries to identify the hot keys (update-wise) and
places them in a separate location in the value log. As a result it
contains the fragmented segments in certain areas of the value log and
then performs GC only these areas. \name{} is orthogonal to HashKV as
it could adopt its techniques for the GC process of the large value
log.  SplitKV~\cite{splitkv} is a single level key value store for
byte addressable NVM and NVMe devices. It performs KV separation for
all KV pairs and place small keys($< 4 KB$) in NVM and large $ \geq
4~KB$ in NVMe. and keeps a global B+-tree index.  When NVM is full, it
transfers data to NVMe. Since all KV pairs eventually end up in NVMe,
it incurs the same GC costs as the previous systems. The approach of
SplitKV is orthogonal to \name{}. \name{} could manage the NVMe device
log more efficiently, while SplitKV manages the NVM layer.

Another prominent technique that reduces I/O amplification is
tiering~\cite{stepmerge}.  In a tiered organization each level
contains a set of sorted runs which contain overlapping key
ranges. Tiering in contrast with leveling reduces
amplification~\cite{sifrdb}
with the penalty of expensive reads since you must check each sorted
run per level. Systems such as Jungle~\cite{jungle},
SplinterDB~\cite{splinterdb},EvenDB~\cite{evendb}, and
PebblesDB~\cite{pebblesdb} use forms of this technique.
\name{} tries to reduce I/O amplification for leveled LSM KV stores
through hybrid KV placement. 



\section{Conclusions}
\label{sec:conc}

In this paper, we design \name{}, a persistent LSM key value store for
fast storage devices. \name{} first models and analyzes the benefits of
using a log for KV separation. Based on this analysis it performs
hybrid placement of KV pairs to reduce I/O amplification and improve
space management.  It does this by keeping small KV pairs in place,
medium KV pairs in logs until the last level(s), and by always using a
log for large KV pairs. Compared to RocksDB, \name{} increases CPU
efficiency by up to 18.7x, decreases I/O amplification by up to 27.1x
at the expense of increasing randomness of I/Os.
We believe that \name{} techniques are compatible with
production systems such as RocksDB and BlobDB, which can  adopt them to reduce 
I/O amplification and increase CPU efficiency.



\bibliographystyle{plain}
\bibliography{paper}
\end{document}